%% file: hidden_state.tex
\pdfoutput=1
\documentclass[11pt]{article}

\usepackage[utf8]{inputenc}
\usepackage[T1]{fontenc}
\usepackage{lmodern}
\usepackage[margin=1in]{geometry}
\usepackage{authblk}
\usepackage{graphicx}
\usepackage{booktabs}
\usepackage[section]{placeins}
\usepackage{amsmath, amssymb}
\usepackage{xcolor}
\usepackage[colorlinks=true, allcolors=blue]{hyperref}
\usepackage[capitalise]{cleveref}
\usepackage[round, sort&compress]{natbib}

\input{numbers}

\title{Crouching Task, Hidden State:
Why Condition Decoding Does Not Establish State Decoding}

\author[1]{Tom Touati}
\author[1]{Tomer Fekete\thanks{Correspondence: \href{mailto:tomer.fekete@brain.space}{tomer.fekete@brain.space}}}
\affil[1]{brain.space}

\date{}

\begin{document}

\maketitle

\input{sections/abstract}

\section{Introduction}
\input{sections/introduction}

\section{Data}
\input{sections/data}

\section{Methods}
\input{sections/methods}

\section{Results}
\label{sec:results}
\input{sections/results}

\input{sections/figures_tables}

\input{sections/mot_generalization}

\section{Discussion}
\input{sections/discussion}

\clearpage
\section*{Supplementary methods and tables}
\input{sections/supplement}

\clearpage
\section*{Supplementary figures}
\input{sections/supplement_figures}

\bibliographystyle{plainnat}
\bibliography{refs}

\end{document}

%% file: numbers.tex
\newcommand{\NxUsableMIDs}{861}
\newcommand{\NxTrainMIDs}{517}
\newcommand{\NxValMIDs}{171}
\newcommand{\NxTestMIDs}{173}

\newcommand{\ContextSecondsHalf}{7}     %
\newcommand{\SmootherWindow}{15}        %
\newcommand{\SmootherWindowSeconds}{30} %
\newcommand{\HMMStayProbability}{0.978}  %

\newcommand{\TestLogLoss}{0.4107}
\newcommand{\TestAcc}{84.97}          %
\newcommand{\TestMacroF}{84.55}       %

\newcommand{\TestLogLossSmoothedContinuous}{0.3955}
\newcommand{\TestAccSmoothedContinuous}{85.32}    %
\newcommand{\TestMacroFSmoothedContinuous}{85.06} %
\newcommand{\TestAccHMM}{86.18}         %
\newcommand{\TestMacroFHMM}{85.91}      %

\newcommand{\PearsonHMMMed}{0.971}        %

\newcommand{\NxPairedShared}{158}

\newcommand{\WithETPairedAcc}{85.21}
\newcommand{\WithETPairedMacroF}{84.80}
\newcommand{\NoETPairedAcc}{65.04}
\newcommand{\NoETPairedMacroF}{59.85}

\newcommand{\NEyeAvail}{147}
\newcommand{\NMatchedWindows}{64{,}706}
\newcommand{\AccEyeOnly}{90.28}
\newcommand{\FOneEyeOnly}{90.25}
\newcommand{\AccWithETmatched}{90.78}
\newcommand{\FOneWithETmatched}{90.60}
\newcommand{\EtPairedMeanPP}{+0.42}
\newcommand{\EtPairedCILo}{-0.14}
\newcommand{\EtPairedCIHi}{+1.05}
\newcommand{\EtFullBetterN}{72}
\newcommand{\EtEyeOnlyBetterN}{65}
\newcommand{\EtTiesN}{10}

\newcommand{\MMSTBehavBootstrapN}{2000}
\newcommand{\MMSTMatchWindow}{30}
\newcommand{\MMSTRawRTWithET}{-0.53}
\newcommand{\MMSTRawRTNoET}{-0.08}
\newcommand{\MMSTCondMaxAbsDiff}{0.02}

\newcommand{\BehavPracCogG}{0.46}             %
\newcommand{\BehavPracCogP}{3.4\times10^{-8}} %

\newcommand{\MOTNMIDs}{177}
\newcommand{\MOTBinSeconds}{5}
\newcommand{\MOTMinSupport}{40}
\newcommand{\MOTTailSupport}{150}

\newcommand{\MOTPCogLow}{0.367}
\newcommand{\MOTPCogHigh}{0.524}
\newcommand{\MOTAccLow}{0.971}
\newcommand{\MOTAccHigh}{0.856}
\newcommand{\MOTRTLow}{1.39}
\newcommand{\MOTRTHigh}{1.40}

\newcommand{\MOTCorrectSignedRhoWithET}{-0.008}
\newcommand{\MOTCorrectSignedRhoNoET}{-0.048}
\newcommand{\MOTRTSignedRhoWithET}{-0.072}
\newcommand{\MOTRTSignedRhoNoET}{-0.051}

\newcommand{\MOTDRatingAcross}{0.251}
\newcommand{\MOTDRatingWithin}{-0.005}
\newcommand{\MOTDRatingAcrossCI}{[0.189, 0.310]}
\newcommand{\MOTDRatingWithinCI}{[-0.112, 0.094]}
\newcommand{\MOTDCorrectAcross}{-0.135}
\newcommand{\MOTDCorrectWithin}{-0.014}
\newcommand{\MOTDCorrectAcrossCI}{[-0.165, -0.103]}
\newcommand{\MOTDCorrectWithinCI}{[-0.044, 0.016]}
\newcommand{\MOTDRTAcross}{-0.065}
\newcommand{\MOTDRTWithin}{-0.069}
\newcommand{\MOTDRTAcrossCI}{[-0.091, -0.040]}
\newcommand{\MOTDRTWithinCI}{[-0.095, -0.045]}
\newcommand{\MOTCorrectCeilingPct}{30}
\newcommand{\MOTCorrectDistinctVals}{6}
\newcommand{\MOTCorrectTrialsPerRec}{20}

\newcommand{\MOTAurocRestTrack}{0.960}
\newcommand{\MOTAurocRestTrackCI}{[0.945, 0.974]}

\newcommand{\MOTAurocLowHigh}{0.639}
\newcommand{\MOTAurocLowHighCI}{[0.618, 0.660]}
\newcommand{\MOTDLowHigh}{0.552}
\newcommand{\MOTDLowHighCI}{[0.469, 0.636]}
\newcommand{\MOTAurocLowHighNoET}{0.626}
\newcommand{\MOTAurocLowHighNoETCI}{[0.607, 0.645]}

\newcommand{\MOTAurocLowHighAbove}{151}

\newcommand{\MOTAurocN}{177}
\newcommand{\MOTETDeltaAuroc}{0.013}
\newcommand{\MOTETDeltaAurocCI}{[-0.004, 0.031]}
\newcommand{\MOTETDeltaAurocP}{0.17}
\newcommand{\MOTETDeltaD}{0.069}
\newcommand{\MOTETDeltaDCI}{[-0.002, 0.142]}
\newcommand{\MOTETDeltaDP}{0.11}
\newcommand{\MOTETDeltaRaw}{0.053}
\newcommand{\MOTETDeltaRawCI}{[0.034, 0.073]}
\newcommand{\MOTETSpreadWith}{0.299}
\newcommand{\MOTETSpreadNo}{0.215}
\newcommand{\MOTETSpreadDelta}{0.084}
\newcommand{\MOTETSpreadDeltaCI}{[0.075, 0.092]}
\newcommand{\MOTETSpreadWiderN}{170}

%% file: sections/abstract.tex
\begin{abstract}
\noindent
Stress classifiers are commonly trained to predict experimental conditions
and then interpreted as readouts of internal state. But condition decoding
and state decoding are not the same objective: a model can improve
condition classification by exploiting task structure without improving
correspondence with behaviour. We examine
this problem in a multimodal classifier of seven prescribed conditions of
the Mannheim Multicomponent Stress Test (MMST), using behavioural
correspondence, eye-tracking contrasts, and transfer to Multiple Object
Tracking (MOT) as complementary probes. Across these analyses, condition
classification accuracy and behavioural correspondence were dissociable. Removing
eye tracking markedly reduced condition accuracy, yet the information it added
did not generally strengthen within-condition correspondence with
correctness, missed trials, or reaction time. Conversely, where eye
tracking was recorded, a classifier trained on gaze features alone
performed similarly to the full multimodal model in condition accuracy. The behavioural measures
also failed to define a common scalar target. In MMST,
harder cognitive conditions produced faster but less accurate responses.
In MOT, the unchanged readout strongly separated rest from tracking but only
modestly separated prescribed load levels; it followed mean subjective
difficulty across levels but not differences among repetitions of the same
level. Thus
better classification of an external condition did not imply better recovery
of the intended internal construct. Replacing condition labels with
behaviour does not yield a simple solution because behavioural
primitives may themselves encode different processes. Models intended to
decode internal states should therefore separate recovery of the
experimental manipulation from correspondence with the intended construct.
Experimental conditions and prescribed task parameters can remain useful
sources of weak supervision, but neither constitutes ground truth for
internal state. This motivates automatic labelling methods that combine
multiple observable and contextual signals while preserving their temporal
support, disagreement, and uncertainty.
\end{abstract}

%% file: sections/introduction.tex
Stress classifiers are usually trained with labels supplied by
an experiment: rest, challenge, or several prescribed levels of demand.
The assigned condition serves as the supervised target, and models are trained
on EEG, cardiac, electrodermal, or multimodal measurements to predict it
\citep{yao_2024,fu_2022,benchekroun_2022,wen_2020}. Success is then
assessed by agreement with that same assigned condition.
These labels are convenient and reproducible, but they identify what the
experimenter presented rather than what the participant experienced.
A protocol label describes the intervention; it is not an observation of
the participant's resulting state. The same is true of a prescribed task
parameter such as nominal load or difficulty: it specifies what was
manipulated, not the value of a latent cognitive variable in that
participant.
The same condition can evoke different internal states, and different
conditions can evoke overlapping states. Establishing construct validity
therefore requires evidence beyond agreement with a single labelling
method \citep{campbell_fiske_1959}.

This creates a problem for model evaluation. Higher accuracy against the
protocol label is normally treated as evidence of a better stress
readout. Yet accuracy can also improve when a model discovers stimulus
layout, elapsed time, response routine, or gaze behaviour. Such a model
has become better at recognising the experiment without necessarily
becoming better at measuring stress. Eye tracking makes this distinction
especially visible because different tasks often require different gaze
patterns. More generally, predictive models can exploit stable shortcuts
that perform well against a benchmark without recovering the intended
construct \citep{geirhos_shortcut_2020}.

Behaviour appears to offer firmer ground, but it does not supply a
single natural target either. Correctness, missed responses, reaction
time, and subjective effort are observable, whereas stress is latent.
However, these primitives reflect partly different processes and may
move in opposite directions. A faster response can accompany either
improved performance or a speed-accuracy trade-off. Collapsing such
measures into one score can conceal rather than resolve the ambiguity
\citep{heitz_sat_2014}.

Recent work
continues to strengthen the models, datasets, and evaluation protocols for
stress detection, both in critical overviews \citep{mentis_2024} and in new
multimodal systems \citep{perumalsamy_2026}, while still treating agreement with
the protocol label as the definition of a successful readout. The target itself
has largely gone unexamined. We revisit this
classical setup using seven conditions from the Mannheim Multicomponent
Stress Test (MMST) \citep{reinhardt_mmst_2012}. Rather than treat condition
accuracy as the endpoint, we ask whether better condition classification also
improves correspondence with observables that were not used to train the
model.

We examine that relationship in three ways. MMST correctness, missed trials,
and reaction time test whether the prescribed cognitive sequence has a common
behavioural ordering. A separately trained model without eye tracking shows
how gaze information changes condition recovery, and a model trained on eye
tracking alone shows how much of the administered condition gaze recovers by
itself. Finally, unchanged
application to Multiple Object Tracking (MOT) \citep{pylyshyn_mot_1988}
tests whether the classifier readout follows a second prescribed task
sequence, and subjective ratings test whether that ordering contains
information beyond the prescribed task level.

\begin{figure}[t]
  \centering
  \includegraphics[width=\linewidth]{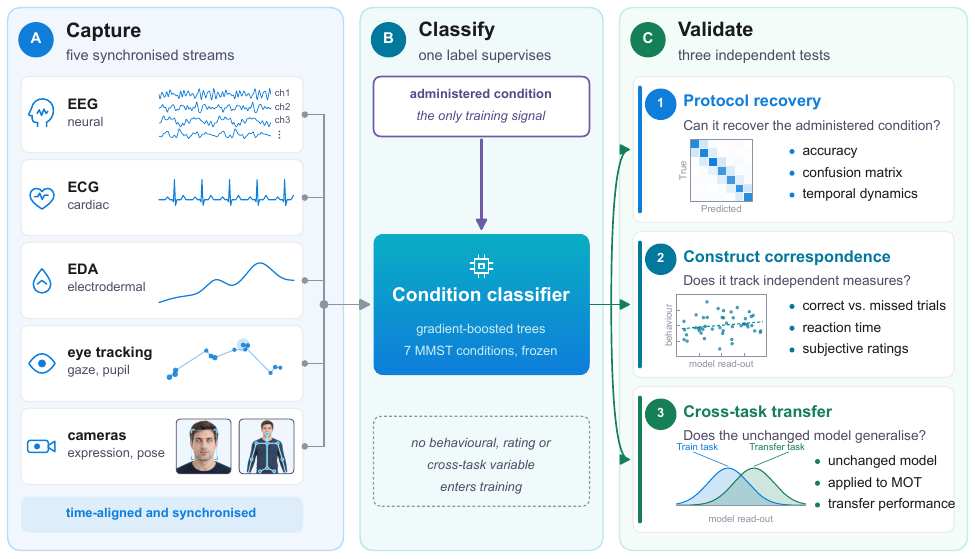}
  \caption{Study design. \textbf{(A)} Five synchronised sensor streams
  were captured and time-aligned; camera video supplied both the face and
  the pose features. \textbf{(B)} They trained a seven-condition
  MMST classifier; only the administered condition labels supervised
  training. \textbf{(C)} Recovery of the administered protocol,
  correspondence with behaviour and ratings, and transfer to MOT were
  evaluated separately.}
  \label{fig:overview}
\end{figure}

Together, these analyses (\Cref{fig:overview}) separate recovery of a labelling rule from evidence
for the intended construct. The distinction applies beyond stress decoding
and motivates automatic labels built from multiple sources of evidence
without prematurely collapsing their dimensions, disagreement, or
uncertainty.

%% file: sections/data.tex
\paragraph{Participants and recordings.}
We selected MMST recordings from the brain.space database for which at
least two sensor modalities were available. Recordings were collected at
four sites and assigned to training, validation, and test sets without
participant overlap, with each site represented in every split. The usable
cohort comprised \NxUsableMIDs\ recordings, assigned \NxTrainMIDs\ to
training, \NxValMIDs\ to validation, and \NxTestMIDs\ to the held-out test
set (demographics and per-modality coverage in
\Cref{tab:cohort_demographics,tab:coverage}).

\paragraph{Protocol.}
Participants completed the Mannheim Multicomponent Stress Test (MMST)
\citep{reinhardt_mmst_2012}. Each session alternated baseline periods with
instruction, practice or affective challenge, and three cognitive blocks
administered at low, medium, and high difficulty. The three levels are
prescribed targets rather than the load each participant met: the cognitive
task was adaptive, setting the pace of the arithmetic to each participant's
performance so that stronger performers received faster, more demanding
sequences, while a monetary reward shown on screen was reduced after errors
and missed responses and was typically depleted by the final block. The
prescribed pace also shortened from the low to the high block, so the levels
order the sessions by intended demand while the difficulty actually
experienced was set per participant. The order of conditions was fixed
across recordings. This fixed order is important: condition, elapsed time,
fatigue, and other cumulative effects cannot be fully separated in the
present data.

\paragraph{Labels.}
Event annotations were grouped into seven conditions: eyes-closed rest,
eyes-open rest, waiting or instruction, practice or affective challenge,
and three cognitive-task levels (Table~\ref{tab:label-map}). These are
protocol labels, not externally established internal states. Behaviour
and self-report were not used as training targets.

\begin{table}[t]
  \centering
  \small
  \caption{The seven conditions used for classification.}
  \label{tab:label-map}
  \begin{tabular}{c l}
    \toprule
    Class & Condition \\
    \midrule
    0 & eyes closed (rest)             \\
    1 & eyes open (rest)               \\
    2 & waiting or instruction         \\
    3 & practice or affective challenge \\
    4 & cognitive task - low           \\
    5 & cognitive task - medium        \\
    6 & cognitive task - high          \\
    \bottomrule
  \end{tabular}
\end{table}

\paragraph{Sensors.}
EEG was acquired with the 115-channel brain.space Sensei headset. Because
high-density EEG would otherwise dominate the multimodal feature space, it
was represented by a reduced 32-channel subset
and a pruned set of spectral features, so that no single modality
overwhelmed the others. These EEG features were combined with ECG,
electrodermal activity, eye tracking, webcam-derived face measures, and
upper-body pose.

\paragraph{MOT transfer cohort.}
The transfer analysis used a separate set of \MOTNMIDs\ Multiple Object
Tracking (MOT) sessions from participants who were absent from the MMST
training and validation splits, so the evaluation is out of subject.
Participants alternated eyes-closed and
eyes-open rest with low- and high-load tracking. Tracking blocks varied
the number of tracked objects, and participants rated difficulty after
each set of trials. MOT condition annotations, behavioural responses,
and ratings were used only to evaluate the unchanged MMST models.

%% file: sections/methods.tex
\paragraph{Signal processing.}
Each modality was processed separately and aligned to the experimental
timeline. EEG preprocessing included high-pass and line-noise filtering,
bad-channel removal, Artifact Subspace Reconstruction \citep{mullen_asr_2015}, and independent
component analysis followed by removal of confirmed ocular and cardiac
components \citep{segal_stst_2026}. The cleaned EEG was represented by
spectral summaries. ECG provided heart-rate and heart-rate-variability
measures, and electrodermal activity was separated into tonic and phasic
components \citep{makowski_neurokit2_2021, boucsein_2012}. Eye-tracker data provided pupil, fixation, saccade, and blink
measures. Face and upper-body movement were estimated from webcam video with MediaPipe
\citep{mediapipe_facelandmarker} and RTMPose \citep{jiang_2023}.

\paragraph{Feature construction.}
Sensor features were placed on a common time grid. Highly redundant and
unstable feature families were removed using the training data. Remaining
features were normalised within each recording with a robust (median and
interquartile) scaler fit on the whole recording rather than on any
dedicated baseline segment. Outlier handling combined a global component,
broad feature bounds estimated once from the training set, with a per
recording refinement in which tighter interquartile bounds were recomputed
from each complete recording.

Each prediction summarised a $\pm\ContextSecondsHalf$ s neighbourhood
around its timestamp. Samples were spaced more densely near the
prediction time and more sparsely at longer offsets.

\paragraph{Classifier.}
We trained a multiclass gradient-boosted tree model
\citep{chen_xgboost_2016}. Model selection used the validation set, and the
test set was used only for final evaluation; the one exception, the
preprocessing recipe of the comparison model without eye tracking, is
described below. Tree interactions were
restricted within sensor families so that a decision could not depend
on an arbitrary conjunction of features from different devices.

\paragraph{Temporal aggregation.}
The putative states such classifiers are intended to track are expected to
persist beyond an individual prediction window. Let $p_{t,c}$ denote the model probability for class
$c$ at prediction window $t$. We therefore also produced a smoothed
probability by averaging the log probabilities over a
\SmootherWindow-window (\SmootherWindowSeconds\,s) centred window and
renormalising,
\begin{equation}
\tilde p_{t,c} = \frac{\exp\!\big(\bar\ell_{t,c}\big)}
{\sum_{c'}\exp\!\big(\bar\ell_{t,c'}\big)},
\qquad
\bar\ell_{t,c} = \frac{1}{|\mathcal{W}_t|}\sum_{s\in\mathcal{W}_t}\log p_{s,c},
\end{equation}
where $\mathcal{W}_t$ is the set of windows in the moving-average
neighbourhood of $t$. This averages the log probabilities and renormalises
to a proper distribution rather than averaging the probabilities directly;
it is a zero-phase step matched to the timescale over which those putative
states are expected to persist. In addition we decoded the most likely class
sequence by Viterbi decoding \citep{forney_viterbi_1973} of a hidden Markov
model \citep{rabiner_hmm_1989} whose emissions are the per-window model
probabilities and whose initial distribution is a class prior $\pi_c$
estimated from the training recordings,
\begin{equation}
\hat s_{1:T} = \operatorname*{arg\,max}_{s_{1:T}}
\left[\log\pi_{s_1} + \sum_{t=1}^{T}\log p_{t,s_t}
+ \sum_{t=2}^{T}\log A_{s_{t-1} s_t}\right].
\end{equation}
The transition matrix $A$ imposed only temporal persistence: a single
self-transition probability of \HMMStayProbability, estimated from the
training recordings, with equal probability assigned to every other
transition. The decoder therefore favours stable sequences without
encoding any particular condition order. Both procedures ran continuously
within each recording and used no annotated condition boundaries; neither
retrains the classifier.

\paragraph{Model without eye tracking.}
To assess the contribution of gaze information we compared the full model
with a separately optimised model that used no eye-tracker features. This
no eye tracking model was not the full pipeline with gaze features simply
removed: it additionally carried rolling-history summaries of the EEG
features, which the full model does not, and it was tuned in its own
hyperparameter search of twenty trials, with its preprocessing recipe
selected using test-set accuracy. The eye-tracking advantage is therefore
a comparison between two separately optimised pipelines rather than a clean
gaze-removal ablation. The comparison was restricted to test recordings
available to both models, and their paired difference quantifies how much
condition classification depends on gaze information in this setting.

\paragraph{Evaluation.}
We report log-loss, accuracy, and macro-F1 on the participant-disjoint
test set, which holds one recording per participant. These metrics are
pooled across all prediction windows in the test recordings. Macro-F1
gives equal weight to each condition despite their unequal durations. The
contribution of eye tracking is quantified as paired recording-level
differences between the models with and without eye tracking. We also
examined the confusion matrix and within-recording agreement between
predicted and prescribed condition sequences.

\paragraph{Behavioural construct checks.}
Reaction time, correctness, and missed trials were not used for model
training. Stress-related responses plausibly unfold across several
seconds and may persist for much longer, whereas an individual action is
momentary. We therefore analyse sustained behavioural summaries rather
than response transients; the optimal choice between full blocks and
shorter within-block intervals remains empirical. For MMST,
the model output and behavioural responses were summarized separately
within practice and cognitive conditions; the analysis did not ask the
model to predict individual responses. The three behavioural measures
were oriented so that higher values indicated better performance,
standardised within participant, and averaged into a composite. Condition
effects were estimated with participant random-intercept models.
Adjacent phases were compared with paired Wilcoxon tests \citep{wilcoxon_1945}
and Holm correction \citep{holm_1979}; the practice-to-low-load effect is
reported as paired Hedges' $g$ \citep{hedges_1981}.

We then compared protocol separation with behavioural correspondence
directly in the \NxPairedShared\ test recordings shared by the models with
and without eye tracking. The readout was the expected class,
$\sum_c c p_c$; as a between-condition anchor we averaged it within practice
and each cognitive condition. To compare readout and behaviour on the same
footing, we placed both at the classifier's \MMSTMatchWindow\,s resolution: the
readout uses the \MMSTMatchWindow\,s smoother, and each behavioural primitive
(correctness, missed trials, reaction time) was taken as a \MMSTMatchWindow\,s
moving average on the same grid. For each recording we computed the Pearson
correlation between readout and behaviour, both raw and after removing the
prescribed-condition mean from each, then averaged across recordings with
\MMSTBehavBootstrapN\ participant-bootstrap draws (resampling recordings,
which is equivalent to resampling participants here because the test split
holds one recording per participant). The correctness and missed-trial
correlations used all \NxPairedShared\ shared recordings; the reaction-time
correlation used the 149 of these with usable reaction-time series. The raw correlation
measures how well the readout tracks behaviour overall; the condition-removed
correlation isolates whether it tracks behaviour beyond the protocol label.
Reaction time is partly mechanical here: the stimulus pacing shortens across
the cognitive blocks, so its dominant variation is set by block position, and
removing the condition mean removes this component; the condition-removed
correlation is therefore the pacing-controlled measure. A smaller share of the
pacing was titrated to each participant, and neither readout tracked this
within-condition component.

For MOT, we asked whether the MMST-derived cognitive score followed a
second prescribed task sequence: rest, low-load tracking, and
high-load tracking. Both MMST models were applied without further training.
The cognitive score is the summed probability assigned to the three MMST
cognitive conditions,
\begin{equation}
s_{\mathrm{cog}} = \sum_{c\in\{\mathrm{CL},\mathrm{CM},\mathrm{CH}\}} p_c,
\end{equation}
where $p_c$ is the model probability for condition $c$ and CL, CM, and CH are
the three cognitive-task conditions of Table~\ref{tab:label-map}. We refer to
this model-derived quantity as the cognitive-class score $s_{\mathrm{cog}}$;
it is not an estimate of a latent cognitive-state probability. It was chosen
because MOT manipulates tracking difficulty. Because a few MOT participants had also
contributed MMST recordings, the transfer cohort was restricted to the
\MOTNMIDs\ sessions from participants absent from the MMST training and
validation splits, so this is an out-of-subject test rather than a
within-subject replay. Tracking accuracy and reaction time described the
behavioural difference between prescribed tracking loads; they were not
classification targets. For the present measure of behavioural correspondence, we
averaged predictions within each high-load tracking block and paired them
with block correctness and reaction time. We then computed
within-participant Spearman correlations and compared them between models,
reporting signed rather than absolute values so that oppositely signed
associations do not cancel into an apparent correspondence. Separately, and
with one estimator for all three independent measures, we asked where any
association lives. Removing each participant's mean from both variables leaves
variation between prescribed levels and gives the across-level estimate;
removing each participant by prescribed level mean instead leaves only
variation among repetitions of the same level and gives the within-level
estimate. Confidence intervals resample recordings with replacement. The
subjective rating uses the four objective sub-levels as the level, the
granularity at which ratings were collected, whereas correctness and reaction
time use the two prescribed tracking loads. This aggregation treats individual responses
as samples of an evolving state without assuming that a block is the
only useful temporal unit. Subjective difficulty was first compared
across prescribed sub-levels. We then centred both rating and model
output within each participant and sub-level to ask the distinct question
of whether the readout tracked differences among repetitions of the same
prescribed difficulty.

\paragraph{Grand average MOT time series.}
To show the transferred readout over the MOT session we aligned every
session on elapsed time, measured in seconds from its first prediction; the
paradigm order is shared, so sessions align without warping. Within each
session we averaged the per-window quantities into \MOTBinSeconds\,s bins,
then averaged those per-session bin values across sessions, reporting the
cross-session mean with a one standard deviation band. The recovered class
is the posterior expectation $\sum_c c\,p_c$, taken directly from the
transferred model's per-window posteriors and shown for the models with and
without eye tracking; the lower panel stacks the cohort mean posterior for
each class from the model with eye tracking, renormalised within each bin.
Prescribed phase bands mark, for each bin, the phase held by the most
sessions, read from the MOT event track and matched to each prediction by
nearest time. Bins were displayed only where at least \MOTMinSupport\
sessions contributed, the elapsed time axis was truncated near the longest
sessions, and a marker flags where fewer than \MOTTailSupport\ sessions
remain. No temporal smoothing or condition alignment was applied; because
the low- and high-load runs alternate in wall-clock time, the later tracking
region pools both levels.

\paragraph{Autonomic profiles.}
As a direct look at the autonomic channels (\Cref{fig:autonomic}) we
summarised heart rate variability, heart rate, and electrodermal level
across the seven administered phases, from eyes-closed and eyes-open rest
and the inter-block waits to practice and the low, medium, and high
cognitive levels. For every recording we placed a common 40\,s window at the
start of each phase, matched in length across phases so the variability
measures are comparable. Practice served as the representative interval
for the combined practice and affective class, and the three inter-block
waiting intervals were averaged within each recording. We re-derived
heart rate variability from the
preprocessed ECG rather than from the model features: R-peak times were
passed through a second beat-correction step, a Kubios-style reconstruction
on a fixed resampling grid followed by a range and relative deviation outlier
rule, implemented with NeuroKit2
\citep{lipponen_hrv_2019, makowski_neurokit2_2021}; the standard deviation of
the corrected inter-beat intervals and the heart rate were then summarised
per window \citep{taskforce_hrv_1996}, with successive difference measures
taken only between adjacent accepted beats.
Electrodermal level was summarised by the mean tonic component in the same
window. Individual windows had to pass beat-quality gates on interval
coverage, per-sample signal quality, and an automated ECG-quality rating,
and recordings whose median heart rate fell outside 50 to 140 beats per
minute were excluded. The resulting complete-case samples were 39
recordings for the heart-derived measures and 167 for electrodermal level.
Phase differences were tested per measure with a one-way within-subject
repeated-measures analysis of variance under the Greenhouse-Geisser
correction \citep{greenhouse_geisser_1959}, with variability measures on a
natural log scale, untransformed means shown, and effect size reported as
generalized $\eta^2$ \citep{bakeman_2005}; pairwise phase comparisons used
paired $t$-tests across all 21 condition pairs with Benjamini-Hochberg
control of the false discovery rate within each measure
\citep{benjamini_hochberg_1995}, summarised as a compact letter display
\citep{piepho_2004}. Error bars are within-subject standard
errors \citep{cousineau_2005, morey_2008}. Labels denote administered phases
and physiological direction only; no stress construct is asserted.

%% file: sections/results.tex
\begin{figure}[tbp]
  \centering
  \includegraphics[width=\linewidth]{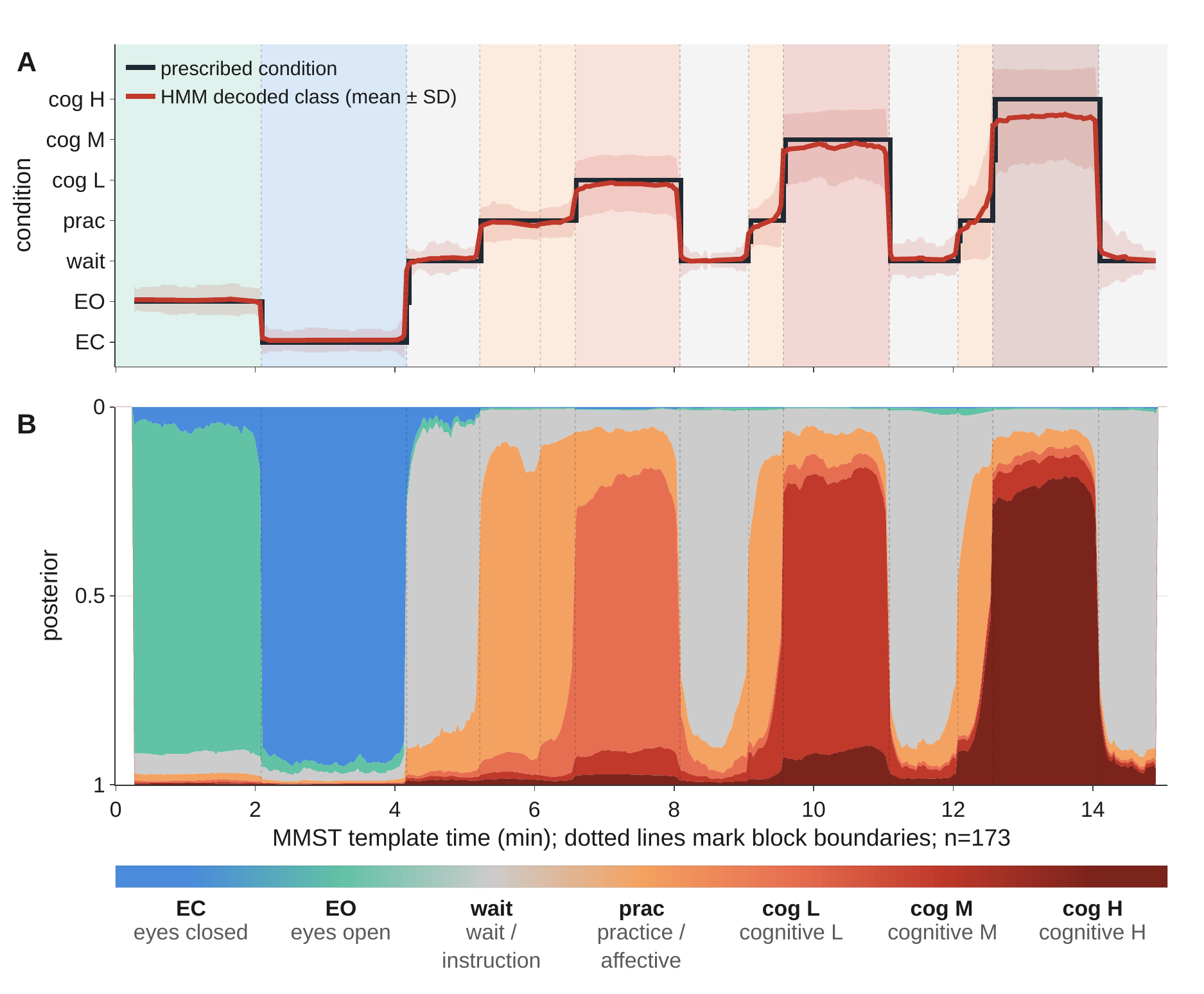}
  \caption{Cohort mean model output across the MMST sequence, on a block
  aligned template time axis (each session's blocks are resampled to a common
  grid before averaging, across the \NxTestMIDs\ held out recordings).
  \textbf{(A)} The prescribed condition (charcoal step) and the cohort mean
  HMM-decoded class (red; Viterbi with a persistence prior, the HMM row of
  Table~\ref{tab:test_metrics} and consonant with \Cref{fig:confusion}),
  superimposed; the band is one standard deviation across recordings and the
  faint colour bands mark the prescribed condition.
  \textbf{(B)} Cohort mean smoothed posterior probability for each
  condition (the classifier's evidence), stacked and inverted so it hangs from the
  top axis. The class key identifies the seven conditions; the colour bands
  align the display to the protocol but are not inputs to the predictions.
  Quantitative metrics are in Table~\ref{tab:test_metrics}; a representative
  single session appears in \Cref{fig:storyboard}.}
  \label{fig:group_timeline}
\end{figure}

\paragraph{The model recovered the prescribed MMST sequence.}
In held-out participants, raw-window classification reached \TestAcc\%
accuracy and \TestMacroF\% macro-F1; temporal aggregation raised accuracy to
\TestAccHMM\% and macro-F1 to \TestMacroFHMM\% with the persistence-based
HMM (Table~\ref{tab:test_metrics}). The recovered sequence closely followed
the prescribed protocol (median within-recording $r=\PearsonHMMMed$;
\Cref{fig:group_timeline}), with errors concentrated around practice and
between neighbouring cognitive levels (\Cref{fig:confusion}).

\begin{figure}[tbp]
  \centering
  \includegraphics[width=\linewidth]{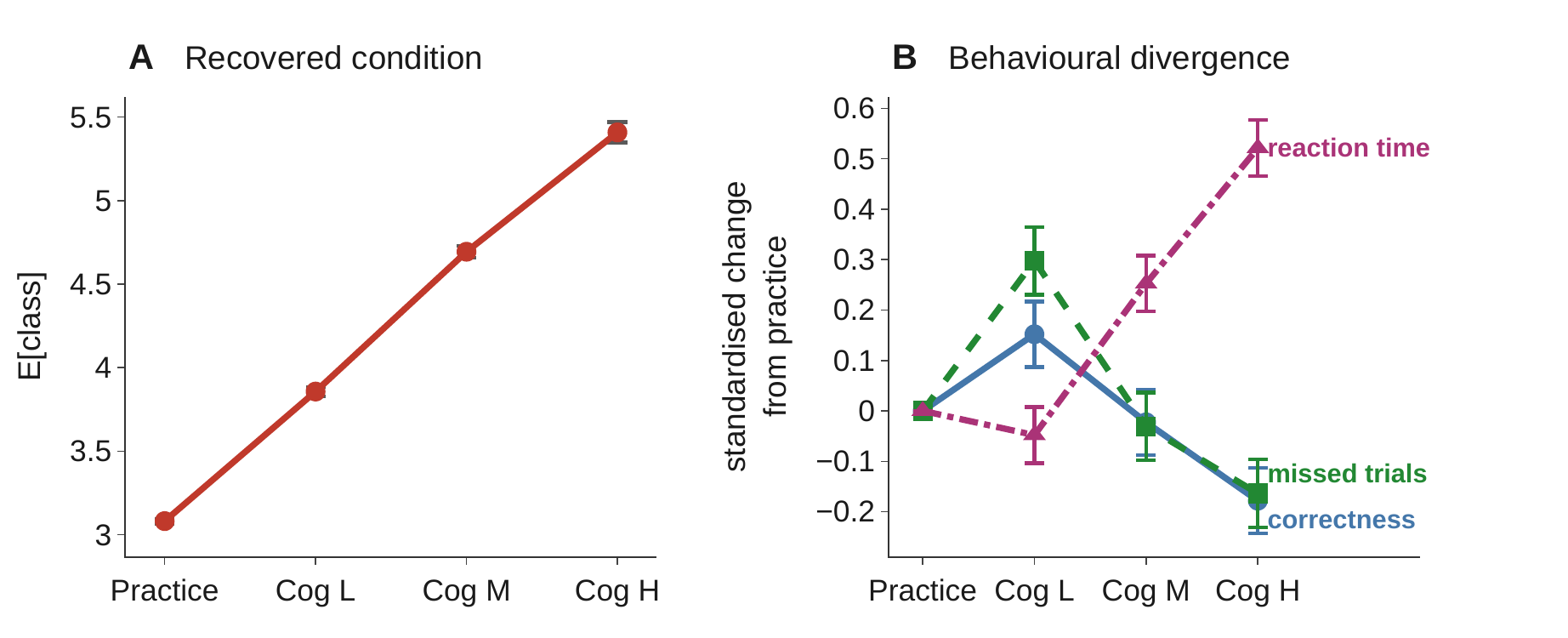}
  \caption{Model output and never-trained measures of MMST performance.
  \textbf{(A)} Mean recovered condition. \textbf{(B)} Participant effects
  relative to practice for correctness, missed trials, and
  reaction time. Higher values indicate better performance. Correctness
  and missed trials worsen beyond low load while reaction times shorten,
  partly reflecting the prescribed acceleration of stimulus pacing, so the
  measures do not define a common load gradient.}
  \label{fig:behav}
\end{figure}

\paragraph{Behaviour does not define the same gradient.}
Performance improved from practice to the low cognitive condition
(Hedges' $g=\BehavPracCogG$, Holm-adjusted $p=\BehavPracCogP$), but did not
continue in one direction across the three cognitive conditions
(\Cref{fig:behav}). Correctness and missed trials worsened beyond low load,
whereas reaction times shortened, partly reflecting the prescribed acceleration
of stimulus pacing. The behavioural measures therefore do not define a common
monotonic gradient.

\begin{figure}[tbp]
  \centering
  \includegraphics[width=\linewidth]{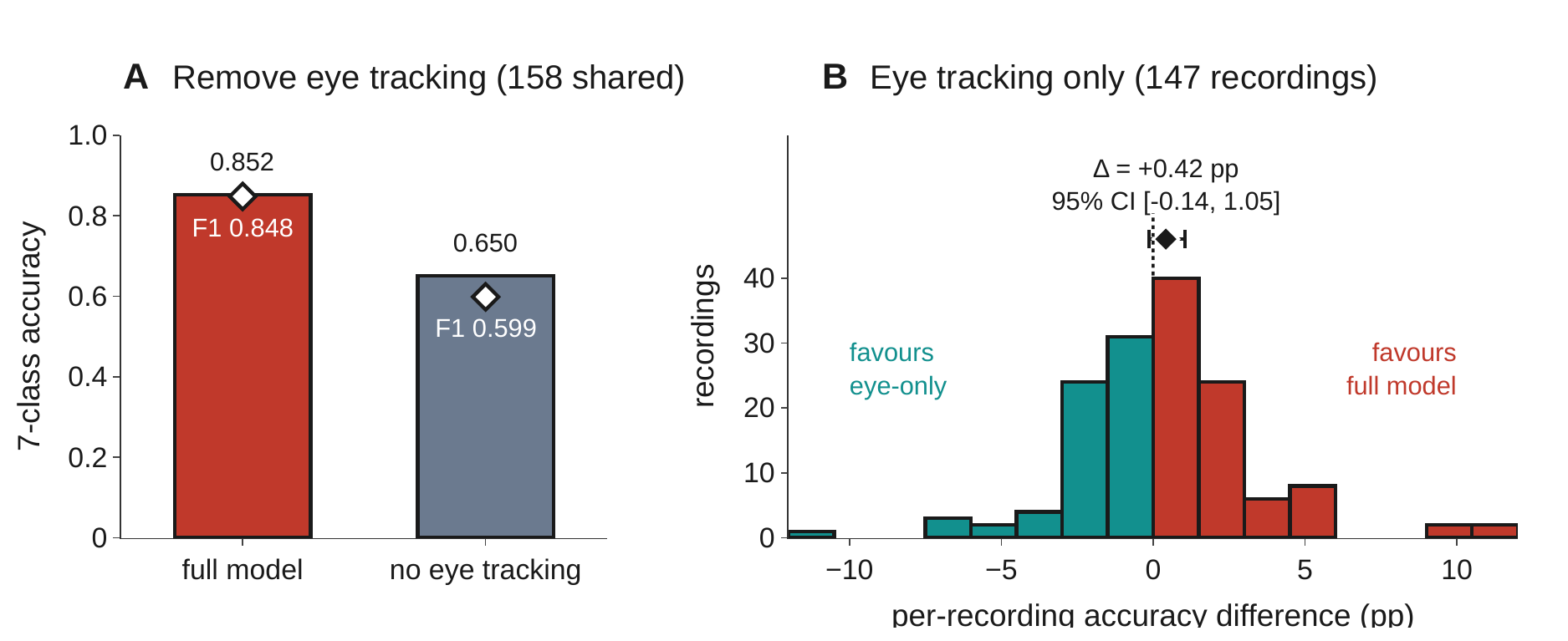}
  \caption{Gaze as a condition classifier.
    \textbf{(A)} On the \NxPairedShared{} shared test recordings, removing eye
    tracking and retraining drops seven-class accuracy from
    \WithETPairedAcc{}\% to \NoETPairedAcc{}\% (bars; overlaid diamond, macro-F1;
    the paired per-recording view is \Cref{fig:noet}).
    \textbf{(B)} A model trained on eye-tracking features alone versus the full
    model, scored on the same wall-clock windows of the \NEyeAvail{} evaluated
    eye-tracking test recordings (\NMatchedWindows{} windows, label agreement
    $1.0$), shown as the per-recording paired accuracy difference (full minus
    eye-only). The distribution straddles zero (mean \EtPairedMeanPP{} pp,
    bootstrap $95\%$ CI $[\EtPairedCILo, \EtPairedCIHi]$; full model ahead on
    \EtFullBetterN{} recordings, eye-only ahead on \EtEyeOnlyBetterN{},
    \EtTiesN{} ties). Bars cover all \NEyeAvail{} recordings; two with
    differences beyond $\pm 12$ pp are shown in the end bins. Red, differences
    favouring the full model; teal, favouring the eye-only model. Where gaze
    is recorded, the administered condition is nearly recoverable from gaze
    alone.}
  \label{fig:etonly}
\end{figure}

\begin{figure}[tbp]
  \centering
  \includegraphics[width=\linewidth]{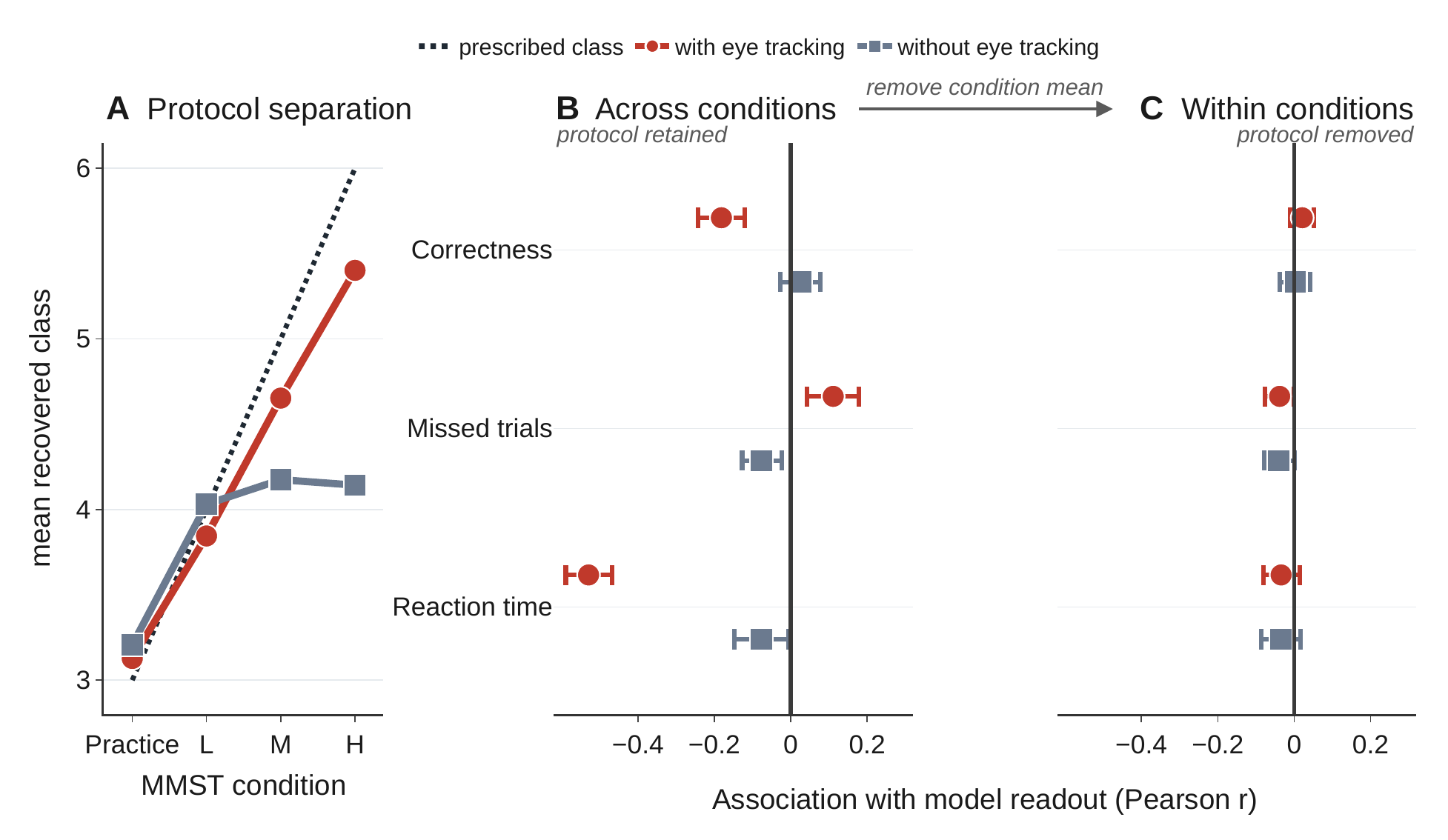}
  \caption{Protocol recovery versus behavioural correspondence in the
  \NxPairedShared\ shared MMST test recordings. \textbf{(A)} Cohort mean
  recovered class across practice and the three cognitive conditions (L, M, H:
  low, medium, and high cognitive load): eye
  tracking preserves separation across the prescribed levels, without it
  the readout flattens after entering the cognitive phase. \textbf{(B)}
  Per-recording correlation between the readout and each behavioural
  primitive at matched \MMSTMatchWindow\,s resolution (raw, protocol
  included): the with eye-tracking readout tracks behaviour markedly more
  than the no eye-tracking readout. \textbf{(C)} The same correlations
  after removing the prescribed-condition mean, which also removes the
  block-fixed stimulus pacing: the eye-tracking advantage disappears and
  neither readout tracks behaviour. Markers are cohort means over
  recordings and bars are participant-bootstrap 95\% confidence intervals.
  Eye tracking sharpens recognition of the administered task, not a read on
  within-task performance.}
  \label{fig:mmst_paradox}
\end{figure}

\paragraph{Eye tracking improves protocol recovery, not within-condition behavioural correspondence.}
Removing eye tracking and retraining reduced accuracy from \WithETPairedAcc\%
to \NoETPairedAcc\% and macro-F1 from \WithETPairedMacroF\% to
\NoETPairedMacroF\% on the \NxPairedShared\ shared test recordings
(\Cref{fig:etonly}A; per-recording paired differences in \Cref{fig:noet}).
With eye tracking, the readout also correlated more strongly with behaviour
at a matched \MMSTMatchWindow\,s resolution, most prominently for reaction
time ($r=\MMSTRawRTWithET$ versus $\MMSTRawRTNoET$; \Cref{fig:mmst_paradox}B).
This advantage disappeared after removing the prescribed-condition mean,
which also removes the block-fixed stimulus pacing: every with minus without
eye-tracking difference in behavioural correlation fell within
$\pm\MMSTCondMaxAbsDiff$ with a confidence interval spanning zero, and
neither readout tracked behaviour within condition
(\Cref{fig:mmst_paradox}C). Gaze therefore greatly improved identification
of the administered task without adding detectable information about how
participants performed within it.

\paragraph{Gaze alone nearly recovers the administered condition.}
A model trained on eye-tracking features alone, compared with the full model
on identical wall-clock windows of the \NEyeAvail\ held-out recordings that
carry eye tracking (\NMatchedWindows\ windows; cohort and matching details
are in the supplementary methods), performed similarly: \AccEyeOnly\% accuracy and
\FOneEyeOnly\% macro-F1 against \AccWithETmatched\% and \FOneWithETmatched\%,
with a per-recording paired difference that straddled zero (mean
\EtPairedMeanPP\ percentage points, participant-bootstrap 95\% interval
$[\EtPairedCILo, \EtPairedCIHi]$; \Cref{fig:etonly}B). The comparison is
confined to eye-available recordings, and an interval that crosses zero is
not an equivalence test; but wherever gaze was recorded, the administered
condition was almost fully recoverable from gaze alone, without EEG,
cardiac, electrodermal, face, or posture input. A high-accuracy classifier of
this stress paradigm can, in this setting, be almost entirely a gaze
classifier.

%% file: sections/figures_tables.tex
\begin{table}[htbp]
  \centering
  \small
  \begin{tabular}{lccc}
    \toprule
    Prediction & Log-loss & Accuracy (\%) & Macro-F1 (\%) \\
    \midrule
    Model output & \TestLogLoss & \TestAcc & \TestMacroF \\
    Continuous \SmootherWindowSeconds\,s average & \TestLogLossSmoothedContinuous & \TestAccSmoothedContinuous & \TestMacroFSmoothedContinuous \\
    HMM (persistence prior) & n/a & \TestAccHMM & \TestMacroFHMM \\
    \bottomrule
  \end{tabular}
  \caption{Performance in held-out participants at the native model
  window and after temporal aggregation. The continuous filter is a
  \SmootherWindowSeconds\,s zero-phase low-pass applied within each
  recording; the HMM decodes the class sequence under a persistence prior
  (self-transition probability \HMMStayProbability, estimated from
  training, with no preferred condition order). Both operate within a
  recording and use no condition boundaries or condition order. The HMM
  produces a hard class sequence, so log-loss does not apply.}
  \label{tab:test_metrics}
\end{table}

%% file: sections/mot_generalization.tex
\section{Transfer of the classifier readout to Multiple Object Tracking}
\label{sec:mot_generalization}

We next applied the unchanged MMST models to \MOTNMIDs\ out-of-subject
Multiple Object Tracking (MOT) sessions \citep{pylyshyn_mot_1988}, in which
participants moved between rest and tracking a subset of moving discs at two
difficulty levels. Because MOT sessions carry no MMST condition labels,
the seven-class MMST classification accuracy is not defined at transfer. We therefore evaluated the
transferred classifier score $s_{\mathrm{cog}}$, defined as the summed model
probability assigned to the three MMST cognitive classes. This is a score derived from the condition
classifier, not a literal MOT label or an independently verified mental state;
it tests whether the condition-derived readout follows a second prescribed task
sequence.

\begin{figure}[!t]
  \centering
  \includegraphics[width=\linewidth]{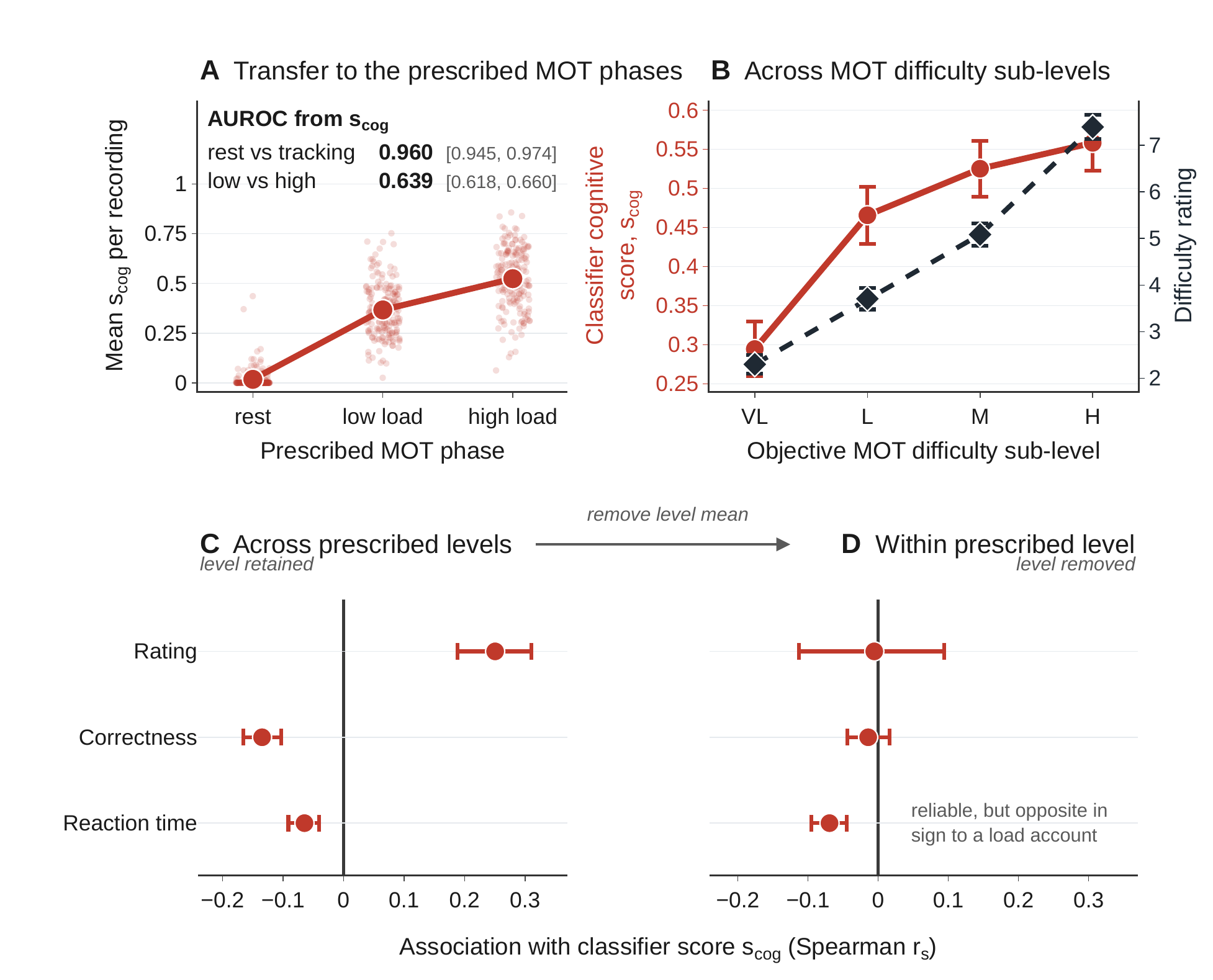}
  \caption{Transfer of the frozen MMST readout to MOT. Because MOT sessions
  carry no MMST condition labels, seven-class classification accuracy is
  undefined at transfer; what transfers is the continuous readout
  $s_{\mathrm{cog}}$, the summed model probability assigned to the three MMST
  cognitive classes.
  \textbf{(A)} Mean $s_{\mathrm{cog}}$ per recording across the three prescribed
  MOT phases, with participant-bootstrap 95\% confidence intervals over the
  faint per-recording points. The inset gives the within-recording AUROC for the
  two contrasts these phases define, rest versus tracking pooling both loads:
  the coarse task distinction transfers almost perfectly, the graded load
  distinction only moderately.
  \textbf{(B)} Across the four objective MOT difficulty sub-levels (VL, L, M
  and H: very low, low, medium and high), $s_{\mathrm{cog}}$ (red, left axis) and
  subjective difficulty (charcoal dashed, right axis) rise together, each in its
  own units, with 95\% confidence intervals. These sub-levels grade the same
  prescribed tracking manipulation that (A), (C) and (D) code coarsely as low
  and high load; difficulty was rated at this finer resolution, so the rating
  rows of (C) and (D) take the sub-level as the level while correctness and
  reaction time take the tracking load.
  \textbf{(C, D)} Association between $s_{\mathrm{cog}}$ and each independent
  measure on one shared scale, computed across prescribed levels (C) and after
  removing each participant's level means (D), with participant-bootstrap 95\%
  confidence intervals. Ratings and correctness are reliable across levels and
  null within one. Reaction time is not: its association is small, reliable in
  both, and runs opposite to a simple load account.}
  \label{fig:mot_behav}
\end{figure}

\paragraph{The frozen MMST readout transferred to MOT task structure.}
MOT does carry prescribed low and high tracking labels, so we measured how well
$s_{\mathrm{cog}}$ discriminated the prescribed MOT contrasts, using the area
under the receiver operating characteristic curve because the readout was never
calibrated or thresholded for MOT. The readout distinguished rest from tracking
with AUROC \MOTAurocRestTrack\ (95\% CI \MOTAurocRestTrackCI), while
discrimination of low- from high-load tracking was more modest but reliable
(AUROC \MOTAurocLowHigh, \MOTAurocLowHighCI; Cohen's $d=\MOTDLowHigh$,
\MOTDLowHighCI; above chance in \MOTAurocLowHighAbove\ of \MOTAurocN\
recordings). Mean $s_{\mathrm{cog}}$ rose from \MOTPCogLow\ at low load to
\MOTPCogHigh\ at high load (\Cref{fig:mot_behav}A), while the participants'
tracking accuracy fell from \MOTAccLow\ to \MOTAccHigh\ and their mean reaction
time was essentially unchanged (\MOTRTLow\ to \MOTRTHigh\,s). Transfer was
therefore strongest for the coarse task distinction and weaker for the
prescribed graded manipulation. Eye tracking did not reliably improve
scale-free low-high discrimination (AUROC difference \MOTETDeltaAuroc, 95\% CI
\MOTETDeltaAurocCI), so the gaze-dependent effect observed in MMST should not be
interpreted as replicated in MOT; the supplementary methods give the three
metrics and why the raw score difference disagrees with them.

\begin{figure}[htbp]
  \centering
  \includegraphics[width=\linewidth]{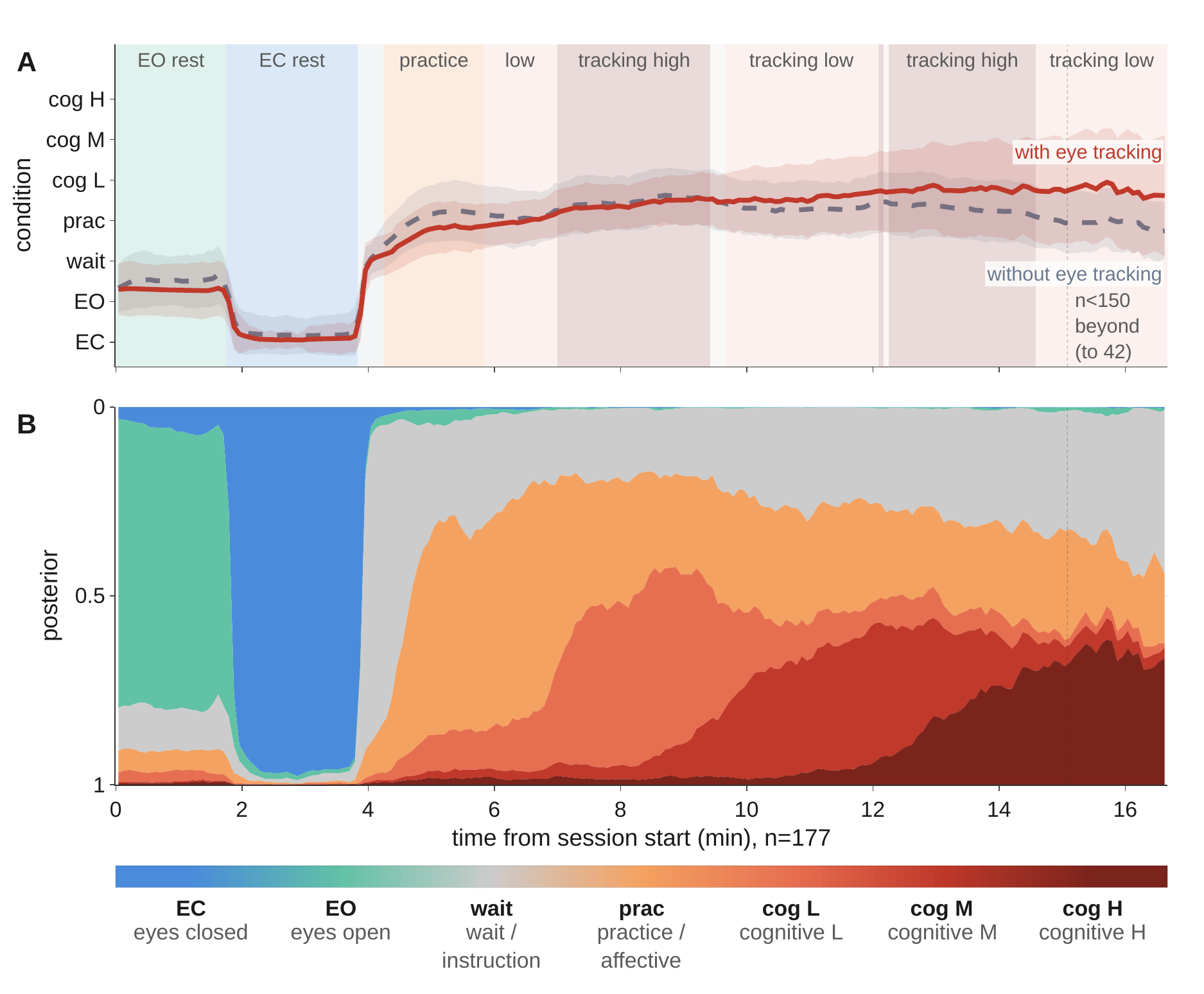}
  \caption{Grand average MOT readout over elapsed time, aligned on seconds
  from session start ($n=\MOTNMIDs$ out-of-subject transfer sessions).
  \textbf{(A)} The cohort
  mean recovered class (the posterior expectation over class index) with eye
  tracking (red, solid) and
  without (slate, dashed), each with a one standard deviation band across
  sessions; faint colour bands and labels mark the cohort-modal prescribed
  phase. \textbf{(B)} The cohort mean posterior for each class with eye
  tracking, stacked and inverted so it hangs from the top axis. The
  class key identifies the seven MMST classes. Eyes-open and
  eyes-closed rest are recovered near the rest classes and the readout rises
  into the cognitive range once the discs move; because the low- and
  high-load runs alternate in wall-clock time the later tracking region
  pools both levels, and the clean low versus high contrast is given in
  \Cref{fig:mot_behav}A. Beyond roughly fifteen minutes fewer than
  \MOTTailSupport{} sessions remain (dotted marker). Annotations align the display to the
  protocol but are not inputs to the model.}
  \label{fig:mot_timeseries}
\end{figure}

\paragraph{Separation across prescribed levels did not extend to within-level correspondence.}
The same ladder continues below the level of the prescribed manipulation.
Across prescribed levels the readout covaried with subjective difficulty
(\Cref{fig:mot_behav}B,C; $r_s=\MOTDRatingAcross$, 95\% CI
$\MOTDRatingAcrossCI$) and with correctness (\Cref{fig:mot_behav}C;
$\MOTDCorrectAcross$, $\MOTDCorrectAcrossCI$), both in the expected direction. Within a level both associations vanished
(\Cref{fig:mot_behav}D; $\MOTDRatingWithin$, $\MOTDRatingWithinCI$; $\MOTDCorrectWithin$,
$\MOTDCorrectWithinCI$). Reaction time behaved differently. Its association was
small but reliable both across levels ($\MOTDRTAcross$, $\MOTDRTAcrossCI$) and
within a level ($\MOTDRTWithin$, $\MOTDRTWithinCI$), and it ran opposite to the
direction a simple load account predicts, in which a higher cognitive score
would accompany slower responses. The same sign appeared in MMST, where the
harder cognitive conditions also produced faster responses. A paired
per-recording comparison within high load gave the same
picture for the model without eye tracking (median signed correlation with
correctness $\MOTCorrectSignedRhoWithET$ with eye tracking and
$\MOTCorrectSignedRhoNoET$ without; with reaction time
$\MOTRTSignedRhoWithET$ and $\MOTRTSignedRhoNoET$), so gaze did not add a correctly
signed within-level correspondence there either. One limit deserves note: correctness at high
load is range restricted, with roughly \MOTCorrectCeilingPct\% of blocks at
ceiling and a median of \MOTCorrectDistinctVals\ distinct values across a
recording's \MOTCorrectTrialsPerRec\ blocks, so its null within-level coupling
should not be read as a strong negative. Read as a sequence, the MOT results
are graded: the readout separates which broad task is being performed almost
perfectly, separates the prescribed load level within that task only
moderately, and provides little consistent information about within-level
variation in the intended construct. The one reliable within-level association,
with reaction time, runs against the direction the task label implies, so it
does not restore a construct-valid ordering. As the question moves from
what the participant was asked to do toward how much of the intended state that
participant experienced, the apparent strength of the readout falls away.
Together, MMST and MOT show that the classifier recovers prescribed task
structure and its correlated signals; they do not establish that it decodes
latent cognitive load or subjective effort.

\FloatBarrier

%% file: sections/discussion.tex
The central result is a divergence between two kinds of success. Within
MMST, eye tracking made the prescribed conditions much easier to classify
and preserved their separation across the cognitive levels, but did not
produce a general increase in within-condition correspondence with
correctness, missed trials, or reaction time. MOT showed the same dissociation from the other
direction: an externally trained readout generalised substantially to the
structure of a different paradigm, nearly perfectly for the coarse task
distinction and moderately for the prescribed load level, yet did not recover a
consistent or construct-valid within-level ordering of behaviour or rated
difficulty. Condition
accuracy and behavioural correspondence were therefore dissociable.

This is not a failure of the classifier. It is a consequence of its
target. Supervised learning rewards any stable signal that predicts the
label. In this experiment, gaze is highly informative because the
conditions differ in their visual layout and required response routines. Adding eye
tracking can therefore produce a better protocol classifier even if the
additional information concerns task identity rather than stress. In these
data the effect is not marginal: on the matched eye-available cohort we
detected no condition-accuracy advantage of the full multimodal model over
eye tracking alone. The
same issue applies, less visibly, to elapsed time, posture, movement, and
other signals that follow the protocol.

The distinction matters because an experimental condition does not
uniquely determine an internal state. Participants can respond
differently to the same challenge, and similar arousal, engagement, or
fatigue can occur in different conditions. The model may recover an
entangled signature of these processes, but accuracy against the
condition label cannot identify the mixture. Improving that accuracy can
even increase reliance on condition-specific cues that are irrelevant to
the intended construct. The same holds for a prescribed task parameter such
as nominal load or difficulty: it records what was manipulated, not the
value the latent variable took in that participant. This is why the MOT
within-level result matters as much as the between-level ordering.

Autonomic measures likewise distinguished rest-like from task-like phases,
even relative to an eyes-open baseline, but did not form a graded sequence
across the administered cognitive levels (\Cref{fig:autonomic}), indicating
genuine physiological engagement without establishing a graded internal
state construct.

MOT provides a useful transfer result, and its shape is informative. The
unchanged MMST score separated rest from tracking almost perfectly, separated
the prescribed load levels only moderately, and did not recover a
construct-valid ordering among repetitions of one level. That is a graded version of the paper's argument: as the
question moves from what task was administered toward how much of the intended
state a particular participant experienced, the apparent strength of the
readout falls away. What transfers is an ordering of the prescribed task
manipulation,
not evidence of a common latent state. Gaze did not reliably sharpen the
transferred load discrimination, so the eye-tracking result remains an MMST
finding rather than one replicated across paradigms.

Behavioural labels are an attractive alternative because they refer to
observable events. Our results also show why that route must remain
multidimensional. Across the MMST cognitive conditions, correctness and
missed trials worsened while reaction times shortened under the changing task
structure. These observables therefore do not supply a common scalar ordering.
Collapsing them into a single performance score would erase that disagreement
and create another ambiguous target. Subjective difficulty poses a related problem: in MOT,
model output and ratings shared a between-level ordering, but the output
did not track rating differences among repetitions of the same level.

Keeping behavioural primitives separate does not require reducing stress
to isolated events. A single correct response or latency is a noisy
expression of an ongoing state, while a full block may conceal meaningful
within-block change. Windows spanning several seconds or shorter
portions of a block offer a plausible middle range, but the useful
temporal resolution is an empirical question. The methodological choice is
therefore not between one scalar condition label and thousands of event
labels, but how to preserve multiple behavioural summaries while testing
the scale at which they relate to the evolving readout.

Classical experimental design addresses these ambiguities by holding
features constant, counterbalancing order, and averaging nuisance
variation to isolate a manipulation. Those controls remain essential for
causal inference, but they solve a different problem from decoding a
state outside the laboratory. Fatigue, learning, adaptation, and
habituation or sensitisation to repeated stressors are not removable
noise in a natural trajectory; they help determine the state that must be
decoded \citep{grissom_habituation_2009}. A nominal low-load condition
presented early in a session need not produce the same state when it
recurs after sustained effort or repeated exposure. Counterbalancing can
estimate an average effect of condition, but it cannot make the condition
label stationary within an individual. The fixed-order MMST preserves one
trajectory of accumulating
effort, fatigue, and adaptation, but does not separate their contributions.
Counterbalanced, gaze-matched manipulations and longitudinal or naturalistic
recordings therefore answer complementary questions: the former can isolate
effects of cognitive demand, while the latter reveal the history that a
real-world decoder must model. The eye-tracking-only benchmark quantifies
how much of the task information gaze carries by itself: where eye tracking
was recorded, we detected no condition-accuracy advantage of the full model
over gaze alone.

Multimodality is consequently more than a route to higher accuracy. It
provides the multiple views needed to characterise a latent state. In these
data, EEG from the 115-channel brain.space Sensei is complemented by cardiac,
electrodermal, gaze, facial, and body pose measurements. Gaze makes task
structure visible, the autonomic signals distinguish broad engagement, EEG
provides a complementary neural view, and behaviour and self-report describe
performance and experience. None is a complete ground truth. Their value lies
partly in testing where the views converge and where they disagree. A model
may fuse many sensor streams yet remain conceptually tied to a single view if
it is trained and judged only against the protocol condition.

The same rationale should extend from model inputs to automatic labelling.
Automatic labelling is then not simply a way to generate more labels; it is
a target-construction problem. Protocol condition, context, behaviour
(actions, accuracy, misses, latency), self-report, physiology, and recent
history provide different observations of an evolving latent state, each
with its own temporal support and uncertainty. Protocol condition can remain
one source of weak supervision among these. Their convergence can support a
higher level label, while disagreement is itself information and should not
be collapsed into the experimental condition or a single prescribed scalar.
Programmatic labelling offers one way to represent noisy and
conflicting supervision sources explicitly \citep{ratner_snorkel_2017}.
Stress is one instance of a wider problem in hidden-state decoding.
In the wild, the target is not a name assigned once from outside, but an
evolving account assembled from multiple, sometimes conflicting views.

%% file: sections/supplement.tex
\paragraph{Model specification.}
\Cref{tab:modelspec} lists the full specification of the two
gradient-boosted tree models \citep{chen_xgboost_2016}: the frozen model
that uses all sensors including eye tracking, whose figures are reported
throughout, and the companion model trained without eye tracking. Both are
seven-class softmax classifiers. They differ in feature composition, in the
number of trees, and in the tuned regularisation, because the two were
tuned separately (see below); the without-eye-tracking model was not the
with-eye-tracking model re-fit on a reduced column set.

\begin{table}[htbp]
  \centering
  \small
  \begin{tabular}{lll}
    \toprule
    Specification & With eye tracking (frozen) & Without eye tracking \\
    \midrule
    Objective & \texttt{multi:softprob} (7 class) & \texttt{multi:softprob} (7 class) \\
    Training metric & \texttt{mlogloss} & \texttt{mlogloss} \\
    Trees (n\_estimators) & 853 & 2198 \\
    Max depth & 10 & 12 \\
    Learning rate & 0.1395 & 0.0526 \\
    Min child weight & 15 & 34 \\
    L1 penalty (reg\_alpha) & 1.594 & 4.981 \\
    L2 penalty (reg\_lambda) & 0.640 & 0.300 \\
    Min split gain (gamma) & 1.374 & 2.332 \\
    Row subsample & 0.496 & 0.660 \\
    Column subsample (per tree) & 0.329 & 0.458 \\
    Interaction constraints & within sensor family & within sensor family \\
    Random seed & 12 & 12 \\
    Split seed & 42 (subject-disjoint) & 42 (subject-disjoint) \\
    Input feature columns & 6820 & 10520 \\
    \quad EEG features selected & 400 (training only) & 400 (training only) \\
    \quad Non-EEG base features & 141 & 126 (no eye tracker) \\
    Rolling-history families & ECG, EDA, eye tracker, & EEG, ECG, EDA, \\
                             & face, pose & face, pose \\
    \bottomrule
  \end{tabular}
  \caption{Specification of the two seven-class XGBoost models. Column
  counts are the flattened feature matrices after context windowing (see
  \nameref{par:featconstruct}); the per-modality base features are
  expanded by the temporal aggregation and, for the listed families, by a
  rolling-history companion. Hyperparameters are the values stored in each
  frozen bundle. Software: XGBoost ($\ge$3.2.0), scikit-learn (1.8.0),
  SciPy ($\ge$1.15.2), NumPy ($\ge$2.2.5), Optuna ($\ge$4.8.0), NeuroKit2
  (development build), and MediaPipe (0.10.32); the frozen fit predates
  this record and the bundle does not pin exact training-time versions.}
  \label{tab:modelspec}
\end{table}

\paragraph{Feature selection and validation-based model selection.}
Feature selection was fit on the training recordings only. The 400 EEG
spectral features were selected on training data; the redundant EEG
relative-history columns were dropped after a training-set screen found
them substitutable for their bare counterparts, whereas the non-EEG
relative-history columns carried unique signal and were kept. Two feature
patterns (\texttt{\_delta} and \texttt{hands\_visible\_ratio}) were excluded
throughout. For the frozen with-eye-tracking model the hyperparameters were
taken from an Optuna study over the training and validation recordings; the
tree count was then fixed at 853, the converged iteration identified by a
separate early-stopping run (patience 75, cap 2000, best iteration 852) on
the same training and validation data, and the held-out test set was scored
exactly once. The without-eye-tracking model followed a separate procedure:
a preprocessing recipe was chosen from a small fixed-parameter sweep whose
variants were compared on held-out test accuracy, then a wider 20-trial Optuna
search was run over its hyperparameters. Because that recipe selection used the
test set, the without-eye-tracking numbers are reported as a separately
optimised pipeline contrast, not a drop-one causal ablation. The sweep variants
differed from one another by under half a percentage point.

\paragraph{Feature construction.}
\label{par:featconstruct}
Each sensor stream was processed independently and then placed on a common
1\,Hz time grid before feature assembly. EEG was recorded at 500\,Hz and
preprocessed with a first-order 1\,Hz high-pass filter, bandstop line-noise
removal at 50\,Hz and its harmonics (2\,Hz stopband, tenth-order
Butterworth, applied at every harmonic up to the Nyquist frequency),
automated bad-channel removal, Artifact Subspace Reconstruction
\citep{mullen_asr_2015} at a standard-deviation cutoff of 20 with a
0.5\,s calibration window, and independent component analysis
\citep{segal_stst_2026} decomposing 42 components followed by removal of
components identified as cardiac (cardiac influence factor $\ge$ 0.95, or
correlation $\ge$ 0.55) or ocular. Before the decomposition the retained
channels were re-referenced to their common average, and the cleaned signal
carried forward to the spectral stage is this average-referenced EEG. Cleaned EEG
was reduced to spectral summaries by a short-time Fourier transform over
1--70\,Hz using a 5\,s window advanced in 1\,s steps, giving one spectral
feature vector per second, followed by light Gaussian smoothing in time and
frequency. The 400 EEG features carried into the models were selected on
the training set.

The non-EEG modalities were summarised on the same 1\,s grid: ECG gave
heart-rate and heart-rate-variability measures; electrodermal activity was
split into tonic and phasic components; the eye tracker gave pupil,
fixation, saccade, and blink measures; and face and upper-body movement
were estimated from webcam video \citep{mediapipe_facelandmarker,
jiang_2023}. Modalities were aligned to the EEG timeline by truncation to
their common support, and short gaps within an available modality were
filled by forward and backward filling. Features were normalised within
each recording with a robust scaler, and outlier clip limits were estimated
from the training set (a global training-derived bound refined by a
per-recording interquartile rule).

For the modality families listed in \Cref{tab:modelspec} a 30\,s
rolling-history companion feature was added: a rolling z-score over a
30-sample (30\,s) trailing window (minimum 10 samples), with the rolling
standard deviation floored at $10^{-3}$ to prevent blow-up on near-constant
stretches. The with-eye-tracking model applied rolling history to ECG, EDA,
eye tracker, face, and pose but not EEG; the without-eye-tracking model
applied it to EEG, ECG, EDA, face, and pose.

Each prediction was made every 2\,s and summarised a $\pm$7\,s
neighbourhood of its timestamp, sampled geometrically so that time points
near the prediction were dense and distant ones sparse; this non-causal
context window was flattened into the per-prediction feature vector,
producing the column counts in \Cref{tab:modelspec}. When a modality family
was unavailable for a recording its columns remained missing rather than
imputed, and XGBoost's built-in missing-value routing sent those inputs
down a learned default branch at each split, so a recording contributed
through whatever sensors it did carry.

\paragraph{Transition sensitivity.}
Context windows are continuous within a recording and grouped by recording
rather than by condition, so a centred window near a block boundary can
include samples from the adjacent administered class, and the 30\,s
rolling-history companion carries information across a transition. To check
that classification is not carried by such mixed context, we measured
with-eye-tracking test accuracy as a function of the distance between each
2\,s prediction window and the nearest condition transition (a block
boundary; each administered block carries a single class), and recomputed
the headline metrics after excluding transition-adjacent windows
(\Cref{tab:transition}). Accuracy is, if anything, slightly lower for windows
straddling a transition (0.82 within the first 3\,s) and rises to a plateau
near 0.855 beyond about 15\,s. Excluding windows within 7\,s of any
transition (59{,}368 of 77{,}082) left the pooled metrics essentially
unchanged and marginally better, and excluding windows within 15\,s
(42{,}009 windows) did likewise. Classification is therefore not carried by
boundary or mixed-context windows; removing them does not degrade
performance.

\begin{table}[htbp]
  \centering
  \small
  \begin{tabular}{lrrrr}
    \toprule
    Window set & Windows & Accuracy & Macro-F1 & Log-loss \\
    \midrule
    All windows & 77{,}082 & 0.850 & 0.845 & 0.411 \\
    Exclude $\pm$7\,s of a transition & 59{,}368 & 0.854 & 0.849 & 0.402 \\
    Exclude $\pm$15\,s of a transition & 42{,}009 & 0.856 & 0.847 & 0.400 \\
    \bottomrule
  \end{tabular}
  \caption{Pooled test metrics (with eye tracking) after removing prediction
  windows near a condition transition. Metrics are computed on the raw argmax
  over the retained windows, identical to the main-text pooling.}
  \label{tab:transition}
\end{table}

\paragraph{Autonomic measures across all eight channels.}
The autonomic analysis computed eight measures and showed three in the main
text. \Cref{tab:autonomic_omnibus} reports all eight on the common 40\,s
window: five heart-derived measures (RMSSD, SDNN, pNN50, mean heart rate,
CVNN; complete-case $n=39$) and three electrodermal measures (tonic level and
phasic amplitude, $n=167$; skin-conductance-response rate, $n=120$), each with
a seven-level within-subject repeated-measures ANOVA (Greenhouse-Geisser) and
the number of the 21 pairwise contrasts surviving Benjamini-Hochberg
correction. RMSSD, SDNN and pNN50 are shown on a natural-log scale for the
test and reported here on the raw scale for readability. The three measures
shown in the main text (SDNN, mean heart rate, tonic electrodermal level) are
the highest-coverage and most interpretable representative of each autonomic
branch: SDNN is the strongest heart-rate-variability omnibus, mean heart rate
is the canonical rate summary and the channel the beat-quality gate is defined
on, and tonic level is the highest-coverage electrodermal channel. The other
five are within-branch variants (RMSSD, pNN50 and CVNN track SDNN; the phasic
electrodermal measures are the phasic counterparts, the phasic-amplitude
omnibus not significant).

\begin{table}[htbp]
  \centering
  \resizebox{\linewidth}{!}{%
  \begin{tabular}{lrrrrrrrrrr}
    \toprule
    Measure & $n$ & EC & EO & wait & prac & L & M & H & $p_{\mathrm{GG}}$ & sig/21 \\
    \midrule
    RMSSD & 39 & $33.0\pm1.5$ & $34.8\pm1.4$ & $33.5\pm1.0$ & $29.3\pm1.6$ & $29.2\pm1.0$ & $29.6\pm1.0$ & $31.5\pm1.6$ & $0.041$ & 2 \\
    SDNN & 39 & $48.1\pm2.6$ & $52.2\pm2.4$ & $51.3\pm1.5$ & $43.6\pm1.8$ & $41.8\pm1.8$ & $39.5\pm1.8$ & $39.5\pm2.5$ & $1.9\times10^{-5}$ & 9 \\
    pNN50 & 39 & $14.2\pm1.3$ & $15.4\pm1.2$ & $13.4\pm0.9$ & $10.5\pm1.3$ & $10.4\pm0.8$ & $10.5\pm0.8$ & $11.8\pm1.2$ & $0.009$ & 2 \\
    Mean HR & 39 & $79.7\pm0.9$ & $78.6\pm0.6$ & $79.5\pm0.5$ & $80.9\pm0.7$ & $81.6\pm0.6$ & $81.4\pm0.7$ & $79.5\pm0.6$ & $0.019$ & 1 \\
    CVNN & 39 & $0.061\pm0.003$ & $0.067\pm0.003$ & $0.066\pm0.002$ & $0.056\pm0.002$ & $0.054\pm0.002$ & $0.052\pm0.002$ & $0.052\pm0.003$ & $4.4\times10^{-4}$ & 8 \\
    EDA tonic & 167 & $1.26\pm0.06$ & $1.30\pm0.05$ & $1.45\pm0.02$ & $1.46\pm0.06$ & $1.51\pm0.03$ & $1.54\pm0.07$ & $1.45\pm0.06$ & $0.009$ & 6 \\
    EDA phasic & 167 & $0.15\pm0.06$ & $0.24\pm0.05$ & $0.25\pm0.02$ & $0.37\pm0.07$ & $0.33\pm0.04$ & $0.26\pm0.09$ & $0.32\pm0.08$ & $0.248$ & 0 \\
    SCR rate & 120 & $0.53\pm0.07$ & $0.41\pm0.06$ & $0.54\pm0.04$ & $0.80\pm0.07$ & $0.61\pm0.07$ & $0.57\pm0.06$ & $0.57\pm0.07$ & $0.003$ & 3 \\
    \bottomrule
  \end{tabular}%
  }
  \caption{Eight autonomic measures across the seven administered conditions,
  cohort mean $\pm$ within-subject standard error on the raw scale. $p_{\mathrm{GG}}$
  is the Greenhouse-Geisser corrected omnibus $p$ from a seven-level
  within-subject ANOVA; sig/21 counts the pairwise condition contrasts
  surviving Benjamini-Hochberg correction. Heart-derived measures are
  complete-case $n=39$; electrodermal $n=167$ (tonic, phasic) or $n=120$
  (SCR rate).}
  \label{tab:autonomic_omnibus}
\end{table}

\paragraph{Usable cohort.}
\Cref{tab:cohort_demographics} and \Cref{tab:coverage} describe the usable cohort: the
861 recordings that entered training, validation, and test (517 / 171 / 173),
contributed by 855 unique participants. Splits were assigned by participant, so
no participant appears in more than one split: six participants contributed two
recordings each, both in the same split (five in training, one in validation),
and the test split holds one recording per participant. The
cohort is almost entirely male (99.2\%), young, and predominantly
right-handed, recorded mostly at one site. Per-modality coverage is
recomputed on this usable set and differs from the pre-filter assignment
totals.

\begin{table}[htbp]
  \centering
  \small
  \begin{tabular}{ll}
    \toprule
    Characteristic & Usable cohort (861 recordings, 855 participants) \\
    \midrule
    Recordings per split & train 517 / validation 171 / test 173 \\
    Age, years (mean $\pm$ SD) & $26.5 \pm 10.4$ \\
    Age, years (median [IQR]) & 22 [20, 29] \\
    Sex & male 854 (99.2\%), female 7 (0.8\%) \\
    Handedness & right 768 (89.2\%), left 77 (8.9\%), both 16 (1.9\%) \\
    Site & Romania Theta Labs 692 (80.4\%), Office 82 (9.5\%), \\
         & Romania Brain Connect 47 (5.5\%), Moldova 40 (4.6\%) \\
    \bottomrule
  \end{tabular}
  \caption{Demographics and acquisition sites of the usable cohort.}
  \label{tab:cohort_demographics}
\end{table}

\begin{table}[htbp]
  \centering
  \small
  \begin{tabular}{lrrrrr}
    \toprule
    Modality & All ($n$) & All (\%) & Train (/517) & Val (/171) & Test (/173) \\
    \midrule
    EEG & 850 & 98.7 & 509 & 171 & 170 \\
    ECG & 855 & 99.3 & 513 & 170 & 172 \\
    EDA & 799 & 92.8 & 473 & 162 & 164 \\
    Eye tracker & 714 & 82.9 & 424 & 139 & 151 \\
    Face & 662 & 76.9 & 398 & 140 & 124 \\
    Pose & 662 & 76.9 & 398 & 140 & 124 \\
    \bottomrule
  \end{tabular}
  \caption{Per-modality coverage on the usable cohort (recordings carrying
  each modality), overall and by split.}
  \label{tab:coverage}
\end{table}

\paragraph{The on-screen pot is endogenous, not a covariate.}
The MMST awards an on-screen monetary pot that decrements only after an
incorrect or missed trial. It is therefore an outcome of task performance
rather than an independent measure of motivation, and it is collinear with
both correctness and the administered load phase: across 903 MMST
sessions the cohort-mean remaining pot fell monotonically with prescribed
cognitive load, from 0.85 in practice to 0.78, 0.35, and
0.07 across low, medium, and high cognitive conditions
(\Cref{fig:pot_supp}). We therefore did not use the pot as a covariate or as
a motivation regressor; it cannot separate motivation from the load
manipulation it is confounded with.

\paragraph{Why the raw eye-tracking gain in MOT is not evidence of sharper transfer.}
The main text reports that eye tracking did not reliably improve scale-free
discrimination of the prescribed MOT load. Three paired per-recording metrics
were computed on the \MOTAurocN\ transfer sessions, and they disagree in an
instructive way. The raw difference in $s_{\mathrm{cog}}$ between high and low
load is larger for the model with eye tracking (\MOTETDeltaRaw, 95\% CI
\MOTETDeltaRawCI). The standardised difference is not
(Cohen's $d$ difference \MOTETDeltaD, \MOTETDeltaDCI, $p=\MOTETDeltaDP$), and
neither is the rank-based one (AUROC difference \MOTETDeltaAuroc,
\MOTETDeltaAurocCI, $p=\MOTETDeltaAurocP$). The reason is scale. The
with-eye-tracking model's within-condition standard deviation of
$s_{\mathrm{cog}}$ is \MOTETSpreadWith\ against \MOTETSpreadNo\ without it, a
paired difference of \MOTETSpreadDelta\ (\MOTETSpreadDeltaCI) that is wider in
\MOTETSpreadWiderN\ of \MOTAurocN\ recordings. A model whose scores occupy a
wider range produces a larger raw condition difference without discriminating
the conditions any better. We therefore report the scale-free metrics and do
not treat the raw gain as evidence that gaze sharpens transferred load
discrimination. Discrimination itself is unaffected by this: AUROC for low
versus high is \MOTAurocLowHigh\ (\MOTAurocLowHighCI) with eye tracking and
\MOTAurocLowHighNoET\ (\MOTAurocLowHighNoETCI) without.

\paragraph{Eye-tracking-only benchmark: cohort and matching.}
\label{par:etonly}
The main text reports a model trained on eye-tracking features alone against
the full model (\Cref{fig:etonly}). A complementary question to the removal
contrast is how much eye tracking carries on its own. The eye-only model reused
the full model's processing code, interaction constraints, rolling-z
companions, and fixed hyperparameters. It is a matched-population benchmark
rather than a literal drop-one ablation: the eye-only model can be trained and scored only
where eye tracking was recorded, and it reads condition labels from the
per-timestamp session log rather than the original event product. Of the $530$
split-assigned training recordings, $425$ carry restored eye tracking; of the
$173$ held-out test recordings, $151$ do, and \NEyeAvail{} of those survived the
eye-only model's windowing and model-preparation stage and were evaluated (the
remaining four were dropped there despite having eye data). We compare the two
models on exactly the same wall-clock windows of those \NEyeAvail{} evaluated
test recordings (\NMatchedWindows{} windows, per-window label agreement $1.0$).

The pooled and paired results are given in the main text and
\Cref{fig:etonly}B: the full model was ahead on \EtFullBetterN{} recordings,
the eye-only model on \EtEyeOnlyBetterN{}, with \EtTiesN{} ties. We do not read
this as the other modalities being uninformative, nor as eye tracking being
provably sufficient: an interval that crosses zero is not an equivalence test.
It does mean there was no detectable accuracy difference between the two models
on this population, at the scale the paired estimate and interval make
explicit. The benchmark is silent on the remaining $26$ test recordings the
eye-only model does not evaluate ($22$ without restored eye-tracking data, plus
the four dropped in preprocessing), where an eye-only model produces no
prediction at all.

Two further caveats apply. The comparison is confined to eye-available
recordings, so it speaks to sufficiency where eye tracking is present, not
across the whole cohort. And, as for the frozen models, features are normalised
within each recording using whole-recording statistics, an offline,
transductive choice applied identically to both models, so it does not bias the
comparison but means the benchmark is not a causal, streaming readout.
Together with \Cref{fig:mmst_paradox}, which shows that eye behaviour tracks the
imposed task structure far more than moment-to-moment performance, the benchmark
locates the gap between classifying the condition and measuring the internal
state the condition was meant to induce.

%% file: sections/supplement_figures.tex
\begin{figure}[htbp]
  \centering
  \includegraphics[width=\linewidth]{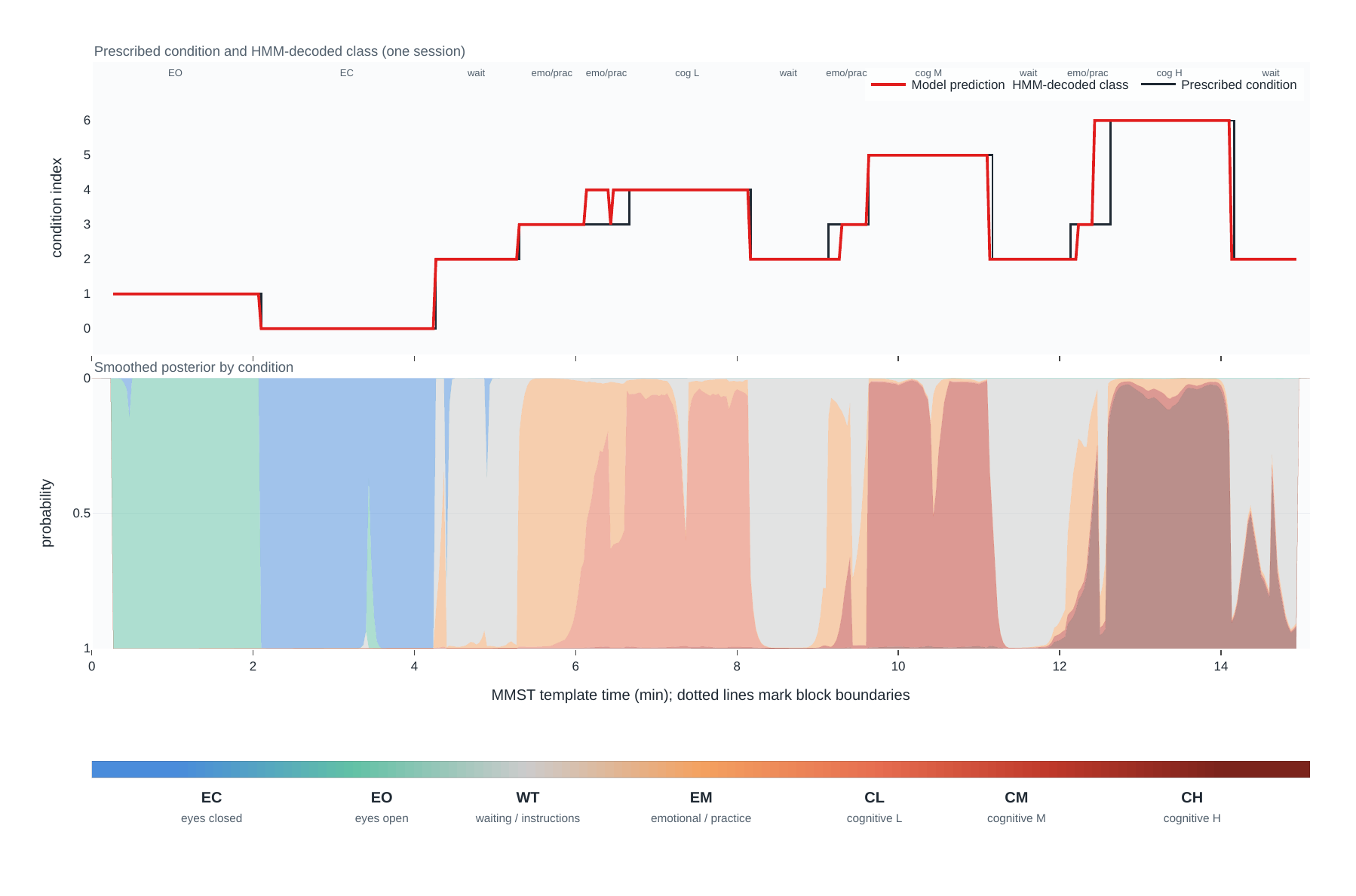}
  \caption{Representative single session (the recording at the cohort
  median of smoothed macro-F1), on a block aligned template timeline.
  \textbf{Top:} prescribed condition (dark step) and the HMM-decoded
  class (red; Viterbi with a persistence prior). \textbf{Bottom:}
  smoothed per-class posterior, inverted so the stack hangs from the top
  axis. The decoded track follows the prescribed sequence within a single
  recording.}
  \label{fig:storyboard}
\end{figure}

\begin{figure}[htbp]
  \centering
  \includegraphics[width=0.85\linewidth]{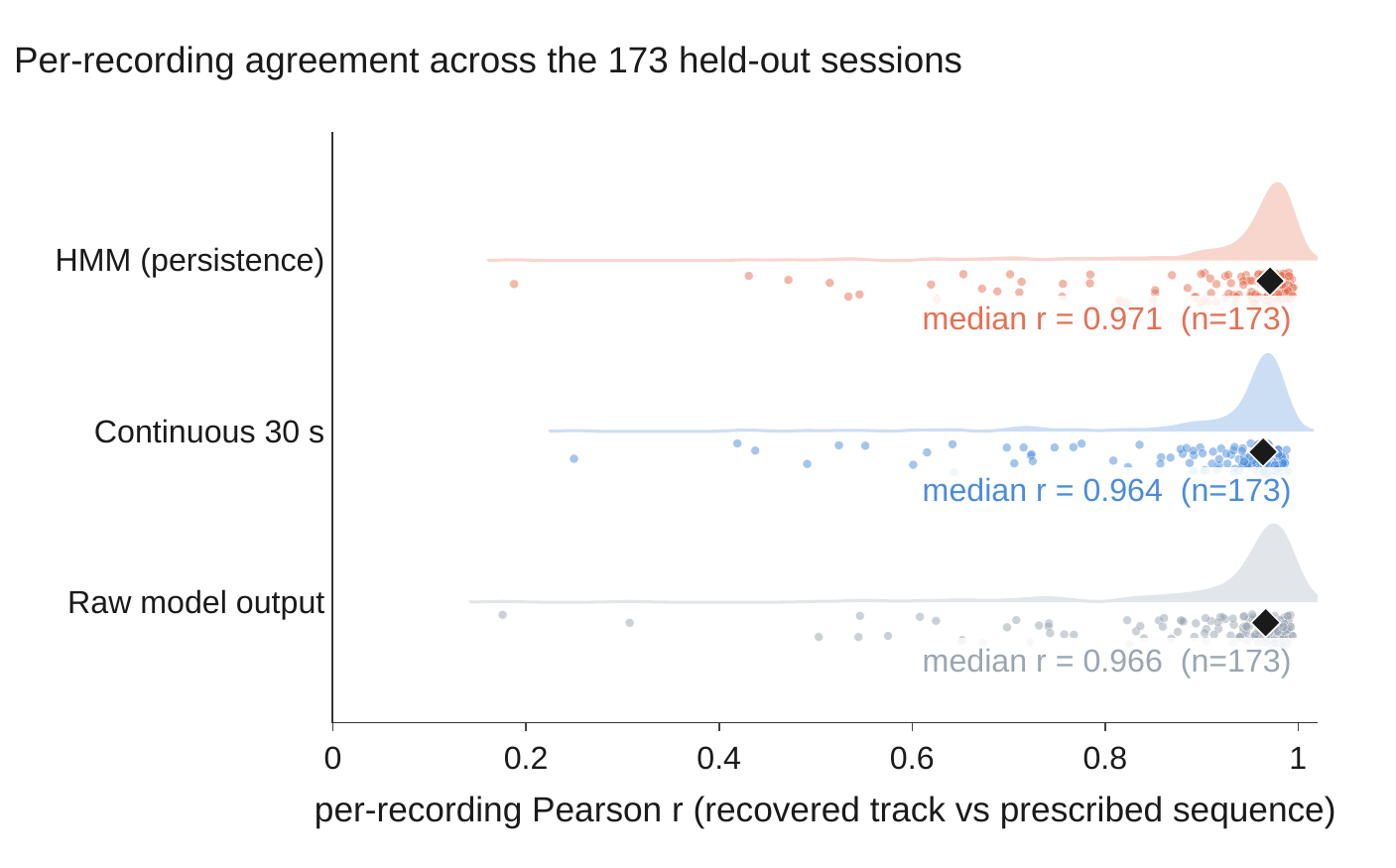}
  \caption{Per-recording agreement between the recovered condition track
  and the prescribed integer sequence, across the \NxTestMIDs\ held-out
  sessions, for the raw, continuous, and HMM readouts. Each row shows the
  distribution of per-recording Pearson $r$ (density and points) with the
  median marked. Agreement is high and tightly concentrated for all three
  readouts.}
  \label{fig:pearson}
\end{figure}

\begin{figure}[htbp]
  \centering
  \includegraphics[width=0.68\linewidth]{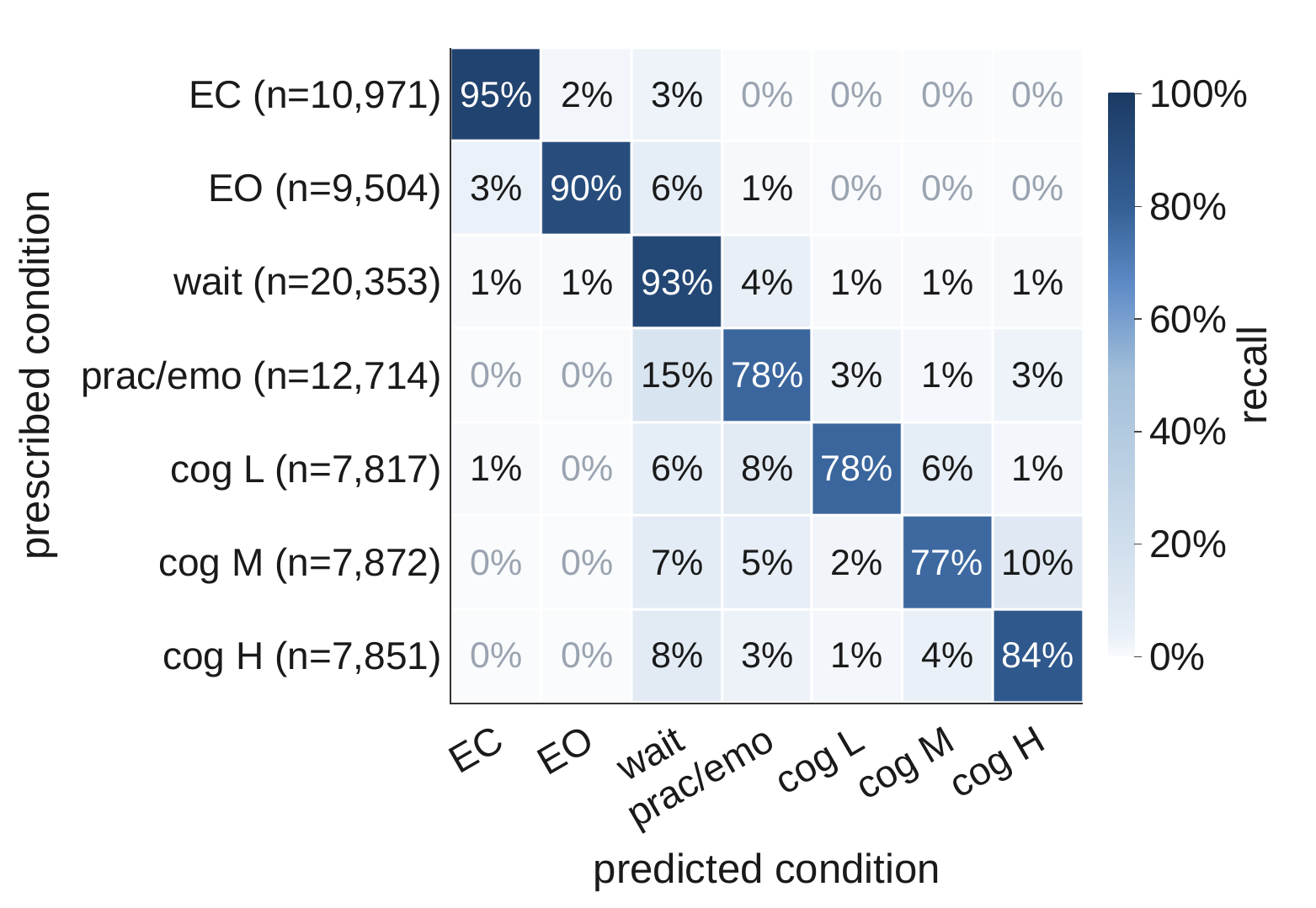}
  \caption{Confusion matrix of the HMM-decoded class on the test set,
  normalised within each prescribed condition (row recall). Row labels give the window support
  $n$ per condition; because each row is normalised, equal cell area
  does not reflect how often a condition occurs. Errors concentrate
  around practice and between neighbouring cognitive conditions.}
  \label{fig:confusion}
\end{figure}

\begin{figure}[htbp]
  \centering
  \includegraphics[width=0.82\linewidth]{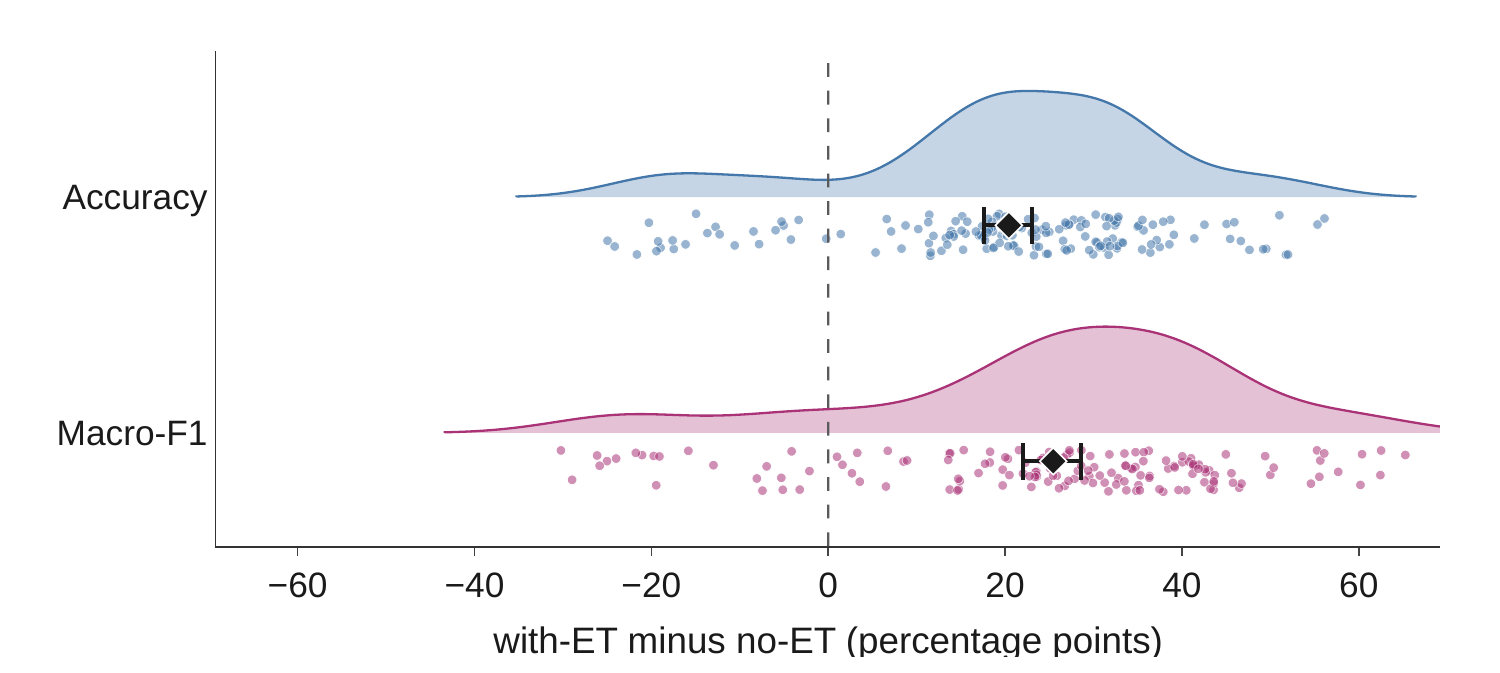}
  \caption{Participant-paired eye-tracker contrast on the \NxPairedShared\ shared test
  recordings (the pooled values are in \Cref{fig:etonly}A). For every recording we take the with eye tracker minus no
  eye tracker difference in accuracy and macro-F1 (percentage points),
  shown as two horizontal rainclouds: the density and the per-recording
  points give the distribution, and the diamond marks the cohort mean
  with a participant-bootstrap 95\% confidence interval. Positive values
  indicate a with eye tracker advantage. The contrast quantifies the
  gaze-dependent protocol information available to the modelling
  pipeline: prescribed conditions elicit distinct visual and response
  routines, and eye tracking exposes those differences.}
  \label{fig:noet}
\end{figure}

\begin{figure}[htbp]
  \centering
  \includegraphics[width=\linewidth]{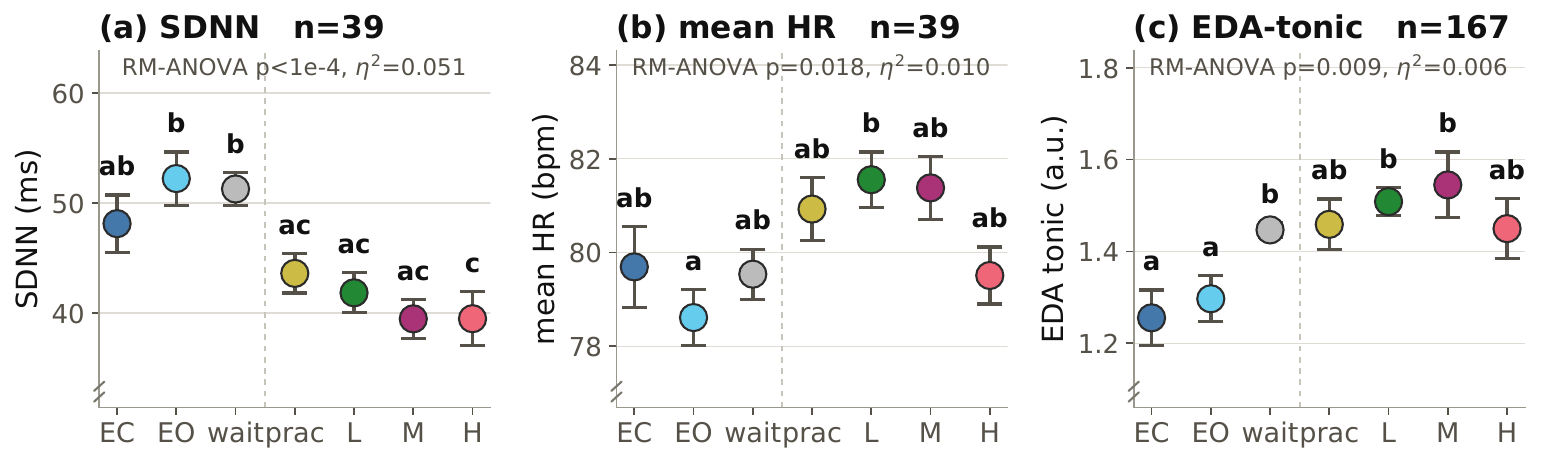}
  \caption{Autonomic engagement across the seven administered phases.
  Within-subject profiles of three autonomic summaries, heart rate
  variability (SDNN), mean heart rate, and tonic electrodermal level, each
  computed on a common short window within every phase and ordered from the
  rest-like phases (eyes-closed, eyes-open, and the waits between blocks) to
  the task-like phases (practice and the low, medium, and high cognitive
  levels). Points are group means with within-subject standard errors on a
  truncated axis (with a break mark), so the small trends remain visible.
  Each panel is annotated with a within-subject repeated-measures ANOVA
  under the Greenhouse-Geisser correction \citep{greenhouse_geisser_1959};
  letters above the points summarise the Benjamini-Hochberg pairwise
  comparisons within each measure \citep{benjamini_hochberg_1995}, and points
  sharing a letter do not differ significantly. Heart rate variability
  (SDNN) and electrodermal measures separate the rest-like from the
  task-like phases but do not form a graded sequence across the cognitive
  levels. Labels denote administered phases and physiological direction
  only; no validated stress construct is asserted.}
  \label{fig:autonomic}
\end{figure}

\begin{figure}[htbp]
  \centering
  \includegraphics[width=0.62\linewidth]{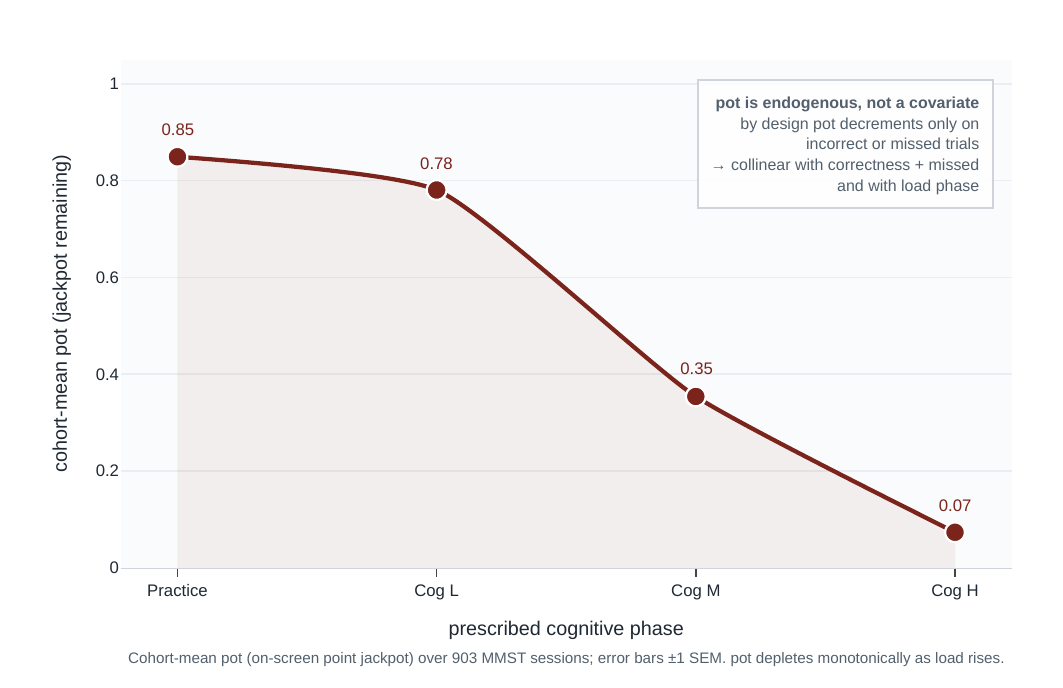}
  \caption{Cohort-mean on-screen pot (jackpot remaining) across the prescribed
  cognitive phases over 903 MMST sessions; error bars are $\pm1$ standard
  error of the mean. The pot decrements only on incorrect or missed trials, so
  it is an outcome of task performance and is collinear with correctness and
  with the administered load phase. The panel is descriptive only.}
  \label{fig:pot_supp}
\end{figure}